\documentclass[sigconf]{acmart}

\AtBeginDocument{%
  }

\usepackage{ifpdf}
\ifpdf
  \pdfoutput=1 
  \usepackage{amsmath}
\else
  \PackageError{main}{This document requires pdflatex. DVI mode not supported.}{Switch to pdflatex and recompile.}
\fi

\usepackage{amsfonts}
\usepackage{bm}
\usepackage[linesnumbered,ruled,commentsnumbered]{algorithm2e}
\usepackage{extarrows}
\usepackage{textcomp}
\usepackage{paralist}
\usepackage{multirow}
\definecolor{brightmaroon}{rgb}{0.76, 0.13, 0.28}
\definecolor{darkgreen}{RGB}{0,100,0}

\newtheorem{definition}{Definition}
\newtheorem{theorem}{Theorem}

\usepackage{enumitem}
\usepackage{pifont}

\hypersetup{
    colorlinks=true,
    linkcolor=blue,
    filecolor=magenta,   
    citecolor=blue,
    urlcolor=blue,
    }

\setcopyright{acmlicensed}
\copyrightyear{2025}
\acmYear{2025}
\setcopyright{acmlicensed}\acmConference[KDD '25]{Proceedings of the 31st ACM SIGKDD Conference on Knowledge Discovery and Data Mining V.2}{August 3--7, 2025}{Toronto, ON, Canada}
\acmBooktitle{Proceedings of the 31st ACM SIGKDD Conference on Knowledge Discovery and Data Mining V.2 (KDD '25), August 3--7, 2025, Toronto, ON, Canada}
\acmDOI{10.1145/3711896.3737146}
\acmISBN{979-8-4007-1454-2/2025/08}

\begin{document}


\title{Taming Recommendation Bias with Causal Intervention on Evolving Personal Popularity}

\author{Shiyin Tan}
\email{tanshiyin1107@gmail.com}
\orcid{0000-0001-8316-2838}
\affiliation{%
  \institution{Institute of Science Tokyo}
  \city{Tokyo}
  \country{Japan}
}

\author{Dongyuan Li}
\authornote{Corresponding author.}
\email{lidy94805@gmail.com}
\orcid{0000-0002-4462-3563}
\affiliation{%
  \institution{The University of Tokyo}
  \city{Tokyo}
  \country{Japan}
}

\author{Renhe Jiang}
\email{jiangrh@csis.u-tokyo.ac.jp}
\orcid{0000-0003-2593-4638}
\affiliation{%
  \institution{The University of Tokyo}
  \city{Tokyo}
  \country{Japan}
}

\author{Zhen Wang}
\email{ zhenwangrs@gmail.com}
\orcid{0009-0005-6450-3699}
\affiliation{%
 \institution{Institute of Science Tokyo}
 \city{Tokyo}
 \country{Japan}}

\author{Xingtong Yu}
\email{ xingtongyu@smu.edu.sg}
\orcid{0000-0002-2884-8578}
\affiliation{%
 \institution{Singapore Management University}
 \city{Singapore}
 \country{Singapore}}

\author{Manabu Okumura}
\email{oku@pi.titech.ac.jp}
\orcid{0009-0001-7730-1536}
\affiliation{%
 \institution{Institute of Science Tokyo}
 \city{Tokyo}
 \country{Japan}}

\begin{abstract}
Popularity bias occurs when popular items are recommended far more frequently than they should be, negatively impacting both user experience and recommendation accuracy. 
Existing debiasing methods mitigate popularity bias often uniformly across all users and only partially consider the time evolution of users or items. However, users have different levels of preference for item popularity, and this preference is evolving over time. 
To address these issues, we propose a novel method called CausalEPP (\underline{Causal} Intervention on
\underline{E}volving \underline{P}ersonal \underline{P}opularity) for taming recommendation bias, which accounts for the evolving personal popularity of users. 
Specifically, we first introduce a metric called {Evolving Personal Popularity} to quantify each user's preference for popular items. 
Then, we design a causal graph that integrates evolving personal popularity into the conformity effect, and apply deconfounded training to mitigate the popularity bias of the causal graph.
During inference, we consider the evolution consistency between users and items to achieve a better recommendation. 
Empirical studies demonstrate that CausalEPP outperforms baseline methods in reducing popularity bias while improving recommendation accuracy.
Our code is available at \textcolor{blue}{\url{https://github.com/ShiyinTan/CausalEPP}}.
\end{abstract}

\ccsdesc[500]{Information systems~Recommender systems}

\keywords{Recommendation, Popularity Bias, Causal Inference}


\maketitle

\section{Introduction}
\label{sec:intro}


\begin{figure}[ht]
\centering
    \includegraphics[scale=0.4]{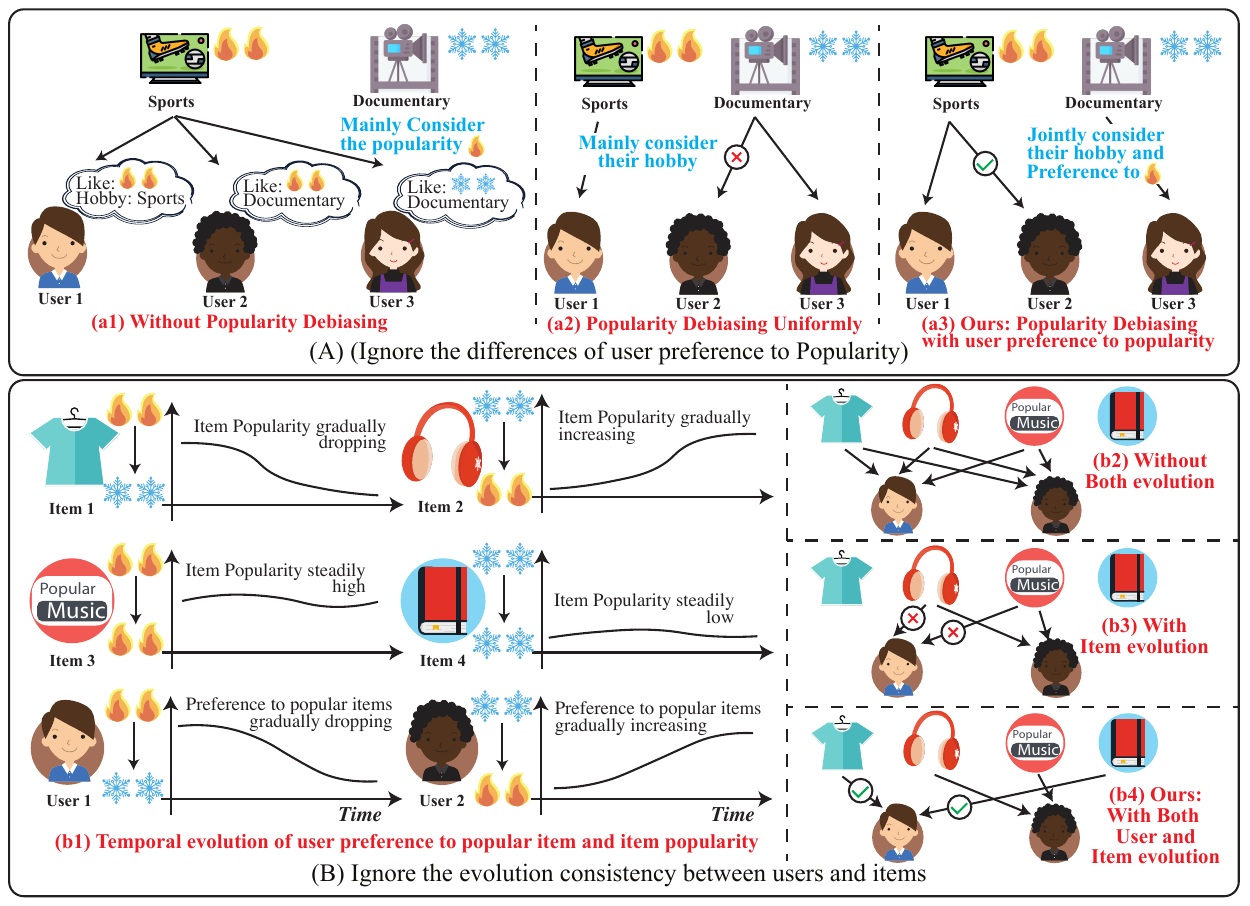}
    \vspace{-15pt}
\caption{(A) Without popularity debiasing, recommender systems recommend popular items to all users. Current work mitigates popularity bias uniformly across all users and ignores the user's preference toward popular items. Our work considers both users' preference for popular items and their hobby. 
(B) Without the evolution of both users and items, recommender systems will recommend items that consider the overall popularity across all times. With only item evolution, they will recommend currently popular items to all users. Our work considers both user and item evolutions, which will consider consistency between the two evolutions. 
}
\label{fig:pre_experiments}
\vspace{-10pt}
\end{figure}


Recommendation systems have been widely deployed to alleviate information overload in various scenarios, including e-Commerce~\cite{DBLP:conf/kdd/WangHZZZL18,DBLP:conf/kdd/0001TS0K24,DBLP:conf/sigir/0003Y00024}, online news~\cite{DBLP:conf/kdd/PanXWYLQQLX024,DBLP:conf/kdd/DuW000024,DBLP:conf/www/GaoFTYCRR24}, and multimedia content~\cite{DBLP:conf/sigir/0001DQZXXXD24,DBLP:conf/sigir/BaiWHCHZHW24,DBLP:conf/www/WeiTXJH24}, which require high-quality user and item representations learned from historical interactions. 
Specifically, they typically analyze user history interactions, such as clicks and purchases, to predict a score for each user-item pair, indicating the likelihood of future interactions. 
Recommender systems can then recommend several items with the highest predicted score to users~\cite{DBLP:conf/uai/RendleFGS09,DBLP:conf/sigir/0001DWLZ020,DBLP:conf/www/ChenLLMX22,YongLi-WWW24}.


Current user-item interaction records in recommendation often exhibit a long-tail distribution~\cite{DBLP:conf/sigir/ZhangF0WSL021}, where a small number of popular items dominate user profiles, while a larger number of less popular items appear infrequently. 
This long-tail distribution of training data arises primarily from the conformity effect, \textit{i.e.,} users frequently interact with items that resemble others' choices, neglecting their own preferences~\cite{DBLP:journals/tkde/ZhaoCZHCZW23,DBLP:journals/fcsc/ChenWCXLH24,DBLP:conf/www/JiangG0CY24}.
Furthermore, classic recommender systems like LightGCN~\cite{DBLP:conf/sigir/0001DWLZ020} and MF~\cite{DBLP:conf/uai/RendleFGS09} often inherit and even amplify this long-tail distribution by over-recommending popular items, known as \textbf{popularity bias}~\cite{DBLP:conf/www/NingC0KHH024,DBLP:journals/corr/abs-2404-12008,DBLP:journals/tois/0007D0F0023}.
Popularity bias gradually narrows user interests and leads to overly homogeneous recommendations~\cite{DBLP:conf/sigir/GeZZPSOZ20,DBLP:conf/kdd/WangF0WC21,DBLP:conf/sigir/BelloginBKLMM24,Zhang_2024,DBLP:conf/www/0003ZWYC23}. To alleviate the popularity bias, numerous debiasing methods have been proposed~\cite{liang2016causal,DBLP:conf/www/NingC0KHH024,DBLP:conf/sigir/0010OMHR24}.
For example, TIDE~\cite{DBLP:journals/tkde/ZhaoCZHCZW23} mitigates the harmful effects of item popularity uniformly across all users. PDA~\cite{DBLP:conf/sigir/ZhangF0WSL021} and DecRS~\cite{DBLP:conf/kdd/WangF0WC21} address popularity bias uniformly as well but also consider the temporal evolution of item popularity during intervention.

Despite their significant success, current popularity debiasing methods still have two key issues under-explored: \textbf{(i) Ignore the differences in user preferences}. 
Previous methods~\cite{DBLP:conf/sigir/NayakGM23,DBLP:conf/sigir/0002WLCZDWSLW22,blundell1988consumer} often address popularity bias uniformly across all users, \textit{i.e.,} recommending unpopular items to users who have no preference for unpopular items based on their hobby, as shown in Figure~\textcolor{blue}{\ref{fig:pre_experiments}(a2)}. 
These approaches overlook individual preferences towards popular items and ignore the balance between preference for popular items and their hobby, resulting in a poor user experience as different users require varying levels of debiasing. 
As shown in Figure~\textcolor{blue}{\ref{fig:pre_experiments}(A)}, users who prefer to click on popular items (users 1 and 2) require a mild level of debiasing and focus more on their preference for popular items. 
In contrast, users who rarely click on popular items (user 3) should undergo a higher level of debiasing and focus more on their hobby.
{Statistical analysis on the Douban-Movie dataset reveals a divide in user behavior: $14975$ users infrequently click on popular items (less than 50\% of clicks on top 10\% popular items), while $32823$ users frequently click on popular items.}
%
\textbf{(ii) Ignore the evolution consistency between users and items}. 
Previous methods often treat item popularity and user preference for popular items as static properties~\cite{DBLP:conf/wsdm/Zhu0ZZWC21,DBLP:conf/kdd/WeiFCWYH21,DBLP:conf/www/NingC0KHH024}.
Alternatively, some approaches focus solely on the evolution of item popularity, simply updating the popularity over time~\cite{DBLP:conf/www/ZhengGCNSJL22,DBLP:conf/kdd/WangF0WC21,DBLP:conf/sigir/ZhangF0WSL021}. 
They both ignore the fact that user decisions are made by the consistency between items and users as time evolves, leading to a mismatch between recommendations and actual user preference.
As shown in Figure~\textcolor{blue}{\ref{fig:pre_experiments}(b3)}, if only considering the item evolution, currently popular items (item 2 and item 3) will be recommended to users who currently have no preference for popular items (user 1), which not consider the evolution of user preference and the consistency between user preference and item popularity. 
{A statistical analysis of the evolution of users’ interactions with popular items classifies users into four groups: $14709$ users shift from frequent to infrequent interaction with popular items; $18459$ users shift from infrequent to frequent; $1065$ users stay infrequent; $14491$ users stay frequent. }

To address the above-mentioned issues, we propose a recommendation debiasing method that makes \underline{Causal} Intervention on \underline{E}volving \underline{P}ersonal \underline{P}opularity (CausalEPP), considering the user's preference for popular items and tracking the temporal evolution of both item popularity and user preference to improve recommendation accuracy. Firstly, to perform different levels of popularity debiasing across different users, we introduce a novel concept called \textbf{evolving personal popularity}, measuring user preference towards popular items by calculating the frequency of a user clicking on high-popularity items. 
Then, we use a \textbf{causal graph} to model the causal effect of recommendation bias. By deconfounded training on this causal graph, we recommend items based on users’ evolving personal popularity and items’ popularity, \textit{i.e.,} items with high popularity will not be recommended to users with low {personal popularity}. 
Furthermore, to mitigate only the harmful part of the popularity bias, we separate item quality from popularity in the causal graph and propose a quality loss to ensure quality is proportional to global popularity (the overall popularity across all time)~\cite{DBLP:journals/tkde/ZhaoCZHCZW23,DBLP:conf/sigir/0010OMHR24}.
Secondly, we consider the \textbf{evolution consistency between item popularity and user preference} for popular items by applying a time-series forecasting-based intervention on local popularity (item popularity in a short period of time) and {evolving personal popularity} during inference. 
Specifically, we apply a moving average method to analyze the evolution tendencies for both personal popularity and item popularity, and use the tendencies to forecast future values as intervention values. 
This forecasting enables the model to more accurately capture users' future preference and item popularity, rather than depending solely on the last observed values. 
Finally, we compare CausalEPP with ten baseline methods across two backbones.
The results indicate that CausalEPP consistently outperforms all state-of-the-art debiasing baselines, with \textbf{improvements of up to 20 4\% in terms of recall} before intervention. In addition, we show the performance gains of the intervention, investigate the effect of three key components, and highlight the benefit of considering temporal evolution in recommendation. 
The main contributions can be summarized as follows.
\begin{itemize}[leftmargin=1em]
\item 
We define a measure of evolving personal popularity, which captures users' preferences toward popular items. This measure allows us to adaptively address popularity bias across users and perform recommendation according to consistency between personal popularity and items' popularity.
\item We construct a complex causal graph to show the recommendation process with item quality, local popularity, and  {evolving personal popularity}. We also theoretically perform deconfounded training to prove the popularity debiasing process. 
\item We propose a time series forecasting approach to capture the evolution consistency of item popularity and  {personal popularity}, which simultaneously predicts future local popularity and evolving personal popularity to enhance recommendation.
\item Experimental results demonstrate that CausalEPP can achieve SOTA performance on three benchmarks. Additional ablation studies demonstrate the effectiveness of each component.
\end{itemize}

\section{Related Work}

\subsection{Popularity Bias in Recommendation System} 
Many successful recommender systems have emerged~\cite{DBLP:journals/www/WuZQWGSQZZLXC24}, including Bert4Rec~\cite{DBLP:conf/cikm/SunLWPLOJ19}, S3-Rec~\cite{DBLP:conf/cikm/ZhouWZZWZWW20}, and P5~\cite{DBLP:conf/recsys/Geng0FGZ22}. 
However, these models often face the challenge of popularity bias~\cite{DBLP:conf/www/YangHXHLL23}, where popular items are recommended disproportionately more often than their actual popularity warrants. This bias arises from two primary factors: the inherent long-tail distribution of {datasets} and the amplification effect of {recommendation systems}. 
Specifically, {\textbf{(i)}} \textbf{Datasets} are generally skewed towards popular items,
where a small fraction of popular items account for most user interactions~\cite{sanders1987pareto}. 
This kind of long-tail distribution~\cite{anderson2006long} is prevalent in recommendation systems.
Similar patterns can be observed in other domains, such as in e-commerce, where a few products dominate sales, and social networks, where a small number of users have millions of followers. 
{\textbf{(ii)}} \textbf{Recommender systems} often inherit and even amplify existing bias by over-recommending popular items, thus popular items are recommended even more frequently than their original popularity exhibited in the dataset. This phenomenon reflects the Matthew effect~\cite{moller2020not}, where \textit{the rich get richer and the poor get poorer}, and makes the collected data in the future more unbalanced.
Ignoring the popularity bias results in several undesirable consequences: \textbf{(i) Narrowed User Interests.} Overexposure to popular items can lead to ``filter bubbles''~\cite{DBLP:conf/www/NguyenHHTK14}, where users' interests become overly skewed and self-reinforcing. 
\textbf{(ii) Reduced Fairness.} Over-recommending popular items reduces the visibility of less popular but potentially high-quality items, leading to unfair outcomes~\cite{DBLP:conf/recsys/AbdollahpouriMB20,DBLP:conf/sigir/ShiZZF024}. This popularity bias not only limits user choice, but also fails to reflect the true quality of recommendation.

\subsection{Debiasing Recommendation}
Recently, causal inference has gained significant attention in recommender systems~\cite{DBLP:journals/tois/0007D0F0023,DBLP:journals/corr/abs-2404-12008,DBLP:conf/kdd/Cai0WBSWZ024}. These methods can be broadly categorized into two types~\cite{luo2024survey}:  
{\textbf{(i)}} \textbf{Propensity score-based methods} adjust the training distribution by re-weighting training samples based on propensity scores within the training loss~\cite{DBLP:conf/icml/SchnabelSSCJ16,DBLP:conf/wsdm/SaitoYNSN20,wang2018deconfounded,DBLP:conf/sigir/LuoW23}. 
For example, IPW~\cite{liang2016causal} assigns weights inversely proportional to item popularity.
However, these methods may overlook how popularity specifically influences each individual interaction. 
{\textbf{(ii)}} \textbf{Causal graph-based methods} explicitly model causal relationships between variables using causal graphs 
\cite{DBLP:journals/tois/GaoZWFHL24,DBLP:conf/sigir/Chen0GWWH24}. 
These strategies are typically categorized into three types according to their causal structures:  
{\textbf{(ii-1)}} \textbf{Collider models} analyze factors that converge on the same outcome~\cite{DBLP:journals/tnn/DingFHLSZ24}. For example, DICE~\cite{DBLP:conf/www/ZhengGLHLJ21} disentangles user interest and conformity using causal embeddings.
TIDE~\cite{DBLP:journals/tkde/ZhaoCZHCZW23} argues that popularity is not always detrimental, separating benign part (item quality) from harmful part of popularity bias. 
{\textbf{(ii-2)}} \textbf{Mediator models} focus on differentiating between the direct and indirect effects of a treatment on the outcome~\cite{DBLP:conf/kdd/Tang0WGXZ0MZ23}. MACR~\cite{DBLP:conf/kdd/WeiFCWYH21} captures indirect dependencies mediated by the popularity of the item and the conformity of the user.
Based on this, PPAC~\cite{DBLP:conf/www/NingC0KHH024} performs personal popularity debiasing, considering popularity bias between different groups of users. 
{\textbf{(ii-3)}} \textbf{Confounder models} employ backdoor adjustment, which blocks the backdoor path by intervening in the treatment variable~\cite{DBLP:journals/tkde/WangLYCWX23}. For example, DecRS~\cite{DBLP:conf/kdd/WangF0WC21} reduces bias amplification by intervening in user representation, while PDA~\cite{DBLP:conf/sigir/ZhangF0WSL021} mitigates the effect of item popularity by intervening on the items.

\subsection{Comparison with Current algorithms}
The following is the differences of our work from current debiasing methods: {\textbf{(i)}} \textbf{Personalized Debiasing Perspective.} PPAC~\cite{DBLP:conf/www/NingC0KHH024} explicitly groups users based on the similarity of the items they have clicked and performs debiasing for each group. In contrast, CausalEPP introduces an evolving personal popularity that implicitly separates users based on their preference for popular items. Furthermore, while PPAC performs popularity debiasing for different user groups, it does not account for the alignment levels between {evolving personal popularity} and items' popularity. 
{\textbf{(ii)}} \textbf{Causal Graph Structure Perspective.} Similar to TIDE~\cite{DBLP:journals/tkde/ZhaoCZHCZW23}, CausalEPP separates item quality from item popularity in the causal graph. However, TIDE focuses solely on a collider structure, whereas CausalEPP addresses more common confounder structure associated with popularity bias. To accurately capture item quality, CausalEPP introduces a quality loss. 
{Furthermore, CausalEPP constructs and theoretically derives a causal graph that accounts for both evolving personal popularity and item popularity. In contrast, TIDE overlooks the fact that conformity is influenced by both user-specific and item-specific factors.}
{\textbf{(iii)}}  \textbf{Temporal Evolution Perspective.} While PDA~\cite{DBLP:conf/sigir/ZhangF0WSL021} and DecRS~\cite{DBLP:conf/kdd/WangF0WC21} consider the temporal evolution of item popularity and perform interventions by searching for the entire value space, CausalEPP searches for intervention values guided by the observed direction. This approach simplifies the algorithm's task of identifying suitable interventions. 
{In addition, we simultaneously address the evolving consistency of item and user popularity, which is non-trivial and novel challenge.}

\section{Preliminaries}
\noindent \textit{\textbf{Notations.}} In this paper, we use capital letter ($U$), lowercase letter (${u}$), and calligraphic letter ($\mathcal{U}$) to represent a variable, its particular value, and its sample space, respectively. 
Let $\mathcal{U} = \{u_1, u_2, \dots, u_n\}$ and $\mathcal{I} = \{i_1, i_2, \dots, i_m\}$ denote the set of users and items, respectively, where $n$ and $m$ are the number of users and items. 
Let $\mathcal{D}$ denote the user behavior data, which is sequentially collected before time $T$, \textit{i.e.,} $\mathcal{D}=\{(u,i,t)|u\in\mathcal{U}, i\in\mathcal{I}, t\leq T\}$, where $(u,i,t)$ denotes user $u$ clicking on item $i$ at timestamp $t$. 
We use $\mathcal{D}_u$ and $\mathcal{D}_i$ to denote the interactions related to user $u$ and item $i$, respectively. $\mathcal{D}^{t_s, t_e}$ denotes all interactions that occur between time intervals [$t_s$, $t_e$]. $|\cdot|$ represents a function that counts the number of elements in a given set, \textit{e.g.,} $|\mathcal{U}|$ is the number of users in the dataset. 

\vspace{3pt}
\noindent \textit{\textbf{Problem Definition}}. From the probabilistic perspective, the target of recommendation is to estimate $P(Y = 1|u, i)$, which denotes the probability that a user $u$ will like an item $i$, where $Y=1$ denotes the click action. 
Traditional recommendation models are trained to learn a scoring function $f(u, i | \Theta)$ from $\mathcal{D}$ with learnable parameters $\Theta$, which captures user preference
\cite{DBLP:conf/kdd/YangDHZXSZ24,DBLP:journals/fcsc/WangLCYTZY25}.

\vspace{3pt}
\noindent \textit{\textbf{Background Knowledge.}} The following types of background knowledge are involved in this paper: global and local popularity, causal graph, and moving average.

\textbf{(i) Global and Local Popularity}. Here, we first introduce the global and local popularity of items. {Global popularity} refers to the total number of interactions an item receives across all times, which is proportional to its average rating~\cite{DBLP:conf/www/NingC0KHH024,DBLP:journals/tkde/ZhaoCZHCZW23}. 
{Local popularity}~\cite{DBLP:conf/sigir/ZhangF0WSL021} of the items can be defined as follows:

\begin{definition}[\textbf{Local Popularity}] 
\vspace{-2pt}
Given an item $i$, local popularity refers to its interactions within a specific recent time frame: 
\begin{equation}
    p_i^t = \frac{|\mathcal{D}_i^{t-w_1,t}|}{|\mathcal{D}^{t-w_1,t}|}, 
    \label{eq:popularity_bias}
\end{equation}
where $\mathcal{D}^{t-w_1,t}$ represents all interactions during the time interval from $t-w_1$ to $t$ and $w_1$ is the window size for the short-term period. 
\vspace{-2pt}
\end{definition}



\begin{figure}[h]
	\centering
        \includegraphics[scale=0.42]{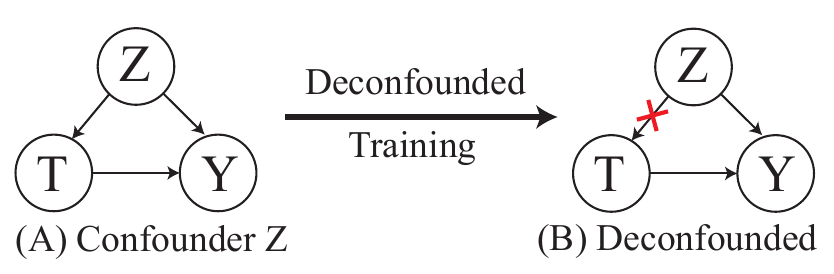}
        \vspace{-5pt}
	\caption{(A) An example of causal graph with confounder. (B) Deconfounded training by eliminating the influence of variable Z on variable T.}
    \label{fig:causal_graph_example}
    \vspace{-8pt}
\end{figure}

\vspace{3pt}

\textbf{(ii) Causal Graph.} As shown in Figure~\textcolor{blue}{\ref{fig:causal_graph_example}}, {causal graphs} provide a straightforward representation of the confounder structure in recommender systems. These directed acyclic graphs describe causal relationships, with nodes representing random variables and edges indicating causal effects between variables. 
%
In Figure~\textcolor{blue}{\ref{fig:causal_graph_example}(A)}, the presence of a confounder $Z$ complicates the analysis of the causal effect from treatment $T$ to outcome $Y$, as $P(Y|T,Z)$. This is because the path $T \rightarrow Y$ includes not only the direct effect of $T$ but also the indirect effect mediated by $Z$. To accurately estimate the causal effect, commonly used \textit{deconfounded training} approach, illustrated in Figure~\textcolor{blue}{\ref{fig:causal_graph_example}(B)}, eliminates the influence of $Z$ by cutting off the path $Z \rightarrow T$ through the application of do-calculus $\text{do}(T=t)$, as $P(Y|do(T=t),Z)$. This intervention isolates the effect of $T$ on $Y$ by simulating a controlled treatment setting. 
Through such deconfounded training, $P(Y|do(T=t),Z)$ estimates outcome $Y$ on treatment $T$ more accurately than $P(Y|T,Z)$.


\vspace{3pt}
\textbf{(iii) Moving Average.}
To measure the trend of evolutions of local popularity and {personal popularity}, we define moving average.
\begin{definition}[\textbf{Moving Average}]
Given time series data with $t$ time points, $\{x^1, x^2, ..., x^t\}$, a moving average is a calculation to analyze data points by creating a series of averages over data points:
\begin{equation}
    \text{MA}_t = \frac{1}{w_2}\sum_{i=t-w_2}^{t}x^{i},
    \label{eq:moving_average}
\end{equation}
where $w_2$ denotes the window size for the average recap and MA can smooth out short-term fluctuations and highlight longer-term trends. 
\end{definition}

\section{Methodology}
In this section, we first introduce {evolving personal popularity} to alleviate the problem of disregarding user individual preference for popular items in \textbf{Section~\textcolor{blue}{\ref{sec:sens_definition}}}. 
Then, we introduce the recommendation process using a causal graph with {evolving personal popularity} and item quality in \textbf{Section~\textcolor{blue}{\ref{sec:causal_graph}}}, which incorporates the {evolving personal popularity} and local popularity into the conformity effect and disentangles the item quality from the popularity bias.
Furthermore, to estimate the popularity bias from train data, we perform the deconfounded training and give the objective function in \textbf{Section~\textcolor{blue}{\ref{sec:training}}}. 
Finally, after estimating popularity bias, we propose a temporal evolution-guided intervention during inference to achieve popularity debasing in \textbf{Section~\textcolor{blue}{\ref{sec:inference}}}. 
The algorithm procedure of CausalEPP is shown in Algorithm~\textcolor{blue}{\ref{alg:link_prediction_algorithm}}


\subsection{Evolving Personal Popularity}
\label{sec:sens_definition}

To consider individual preferences toward popular items, \textit{i.e.,} the more frequently a user interacts with high-popularity items, the more likely the user clicks on popular items in the future, 
we define  {evolving personal popularity} as follows:

\begin{definition}[\textbf{Evolving Personal Popularity}] 
Given a user $u$ at time t, the evolving personal popularity denotes the ratio of clicks on high-popularity items to the total number of user interactions within a time window:
\begin{equation}
    s_u^t = \frac{|\{(u,i):i\in\mathcal{D}_u^{t-w_1,t} \ \mathrm{and}\  p_i^t>\hat{p}^t\}|}{|\mathcal{D}_u^{t-w_1,t}|},
    \label{eq:popularity_sensitivity}
\end{equation}
where $\mathcal{D}_u^{t-w_1,t}$ represents the interactions involving user $u$ between time $t-w_1$ and $t$.
$\hat{p}^t$ is the threshold for high-popularity items. Therefore, $p_i^t>\hat{p}^t$ indicates that item $i$ is popular recently. 
\end{definition}


Since the distribution of local popularity varies over time, the search space for the threshold $\hat{p}^t$ becomes large, since $\hat{p}^t$ may differ at each time point $t$. To address this issue, we set the threshold $\hat{p}^t$ as the top percentile of local popularity, which dynamically defines ``high popularity''. Specifically, we use the top 20\% percentile of popularity values within the window from $t-w_1$ to $t$ as the threshold for high popularity, as shown in Section~\textcolor{blue}{\ref{sec:param_sens}}. This approach limits the search space by adapting the threshold to a recent popularity trend, making the process computationally feasible.

\begin{figure}[ht]
        \vspace{-8pt}
	\centering
        \includegraphics[scale=0.33]{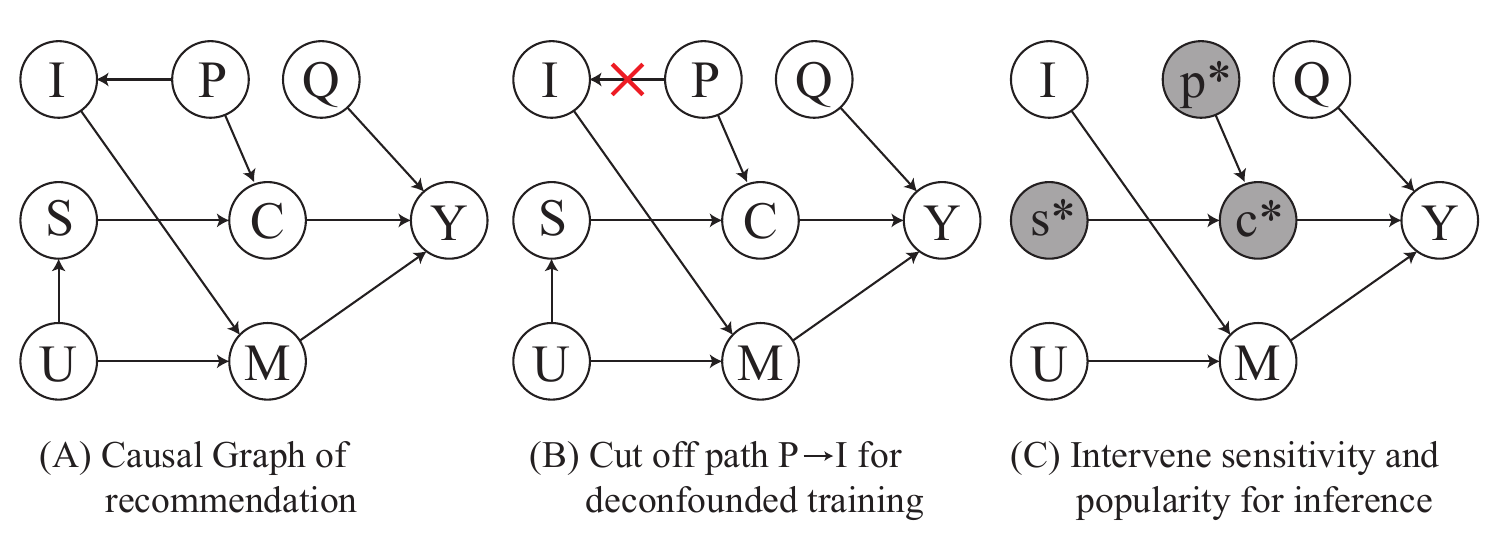}
        \vspace{-3pt}
	\caption{(A): Causal graph to describe the recommendation process that incorporates  {evolving personal popularity} ``S'' into the conformity effect and disentangles quality ``Q'' from popularity ``P''. 
    (B): During training, we cut off the influence from local popularity to items for deconfounded training. (C): During inference, we intervene  {evolving personal popularity} ``$s^*$'' and local popularity ``$p^*$'' for bias adjustment. }
    \label{fig:causal_graph}
    \vspace{-10pt}
\end{figure}

\subsection{Causal Graph}
\label{sec:causal_graph}

Before debiasing, we should use a causal graph to describe the overall recommendation process influenced by all factors. 
As shown in Figure~\textcolor{blue}{\ref{fig:causal_graph}(A)}, our causal graph consists of eight variables: 
\begin{inparaenum}[(1)]
    \item $U$: user;
    \item $I$: item;
    \item $S$:  {evolving personal popularity};
    \item $P$: item's local popularity;
    \item $Q$: item quality;
    \item $C$: the conformity effect;
    \item $M$: the matching score;
    \item $Y$: interaction probability. 
\end{inparaenum}
Next, we explain the edges (\textit{causal effect}) in this causal graph. 
\begin{itemize}[leftmargin=15pt]
    \item $I\leftarrow P$: This edge indicates that local popularity affects the exposure of items, which causes an extremely unbalanced click distribution for popular items after training~\cite{abdollahpouri2020popularity}.
    \item $U\rightarrow S$: This edge represents the  {evolving personal popularity} according to the user's historical behavior.
    \item $Q\rightarrow Y$: This edge represents the impact of the item quality on the user's decision making, as higher-quality items are more likely to be clicked. Quality is separated from popularity. 
    \item $(P,S)\rightarrow C\rightarrow Y$: These edges reflect how the recommendation of the conformity effect is influenced by the  {evolving personal popularity} and the local popularity. 
    \item $(U,I)\rightarrow M\rightarrow Y$: These edges represent the indirect effect of $U$, $I$ on $Y$ through the mediator variable $M$. This path is to develop a statistical model for conditional probability $P(Y|U,I)$.
\end{itemize}

From the causal graph, we can estimate the conditional probability $P(Y|U,I,Q)$. Formally, given $U={u}$, $I={i}$, $Q={q}$, the conditional probability $P(Y|U,I,Q)$ can be derived as follows: 
\begin{subequations}
\begin{align}
&P(Y \mid U = u, I = i, Q = q) \notag\\
&= \frac{\sum_{p \in \mathcal{P}} \sum_{c \in \mathcal{C}} \sum_{m \in \mathcal{M}} \sum_{s \in \mathcal{S}} P(Y, p, c, s, m, u, i, q)}{P(u) P(i) P(q)} \label{eq:correlation_1} \\
&= {\sum_{p,c,m,s} P(p|i) \underbrace{P(c|p,s) P(s|u)}_{\textcolor{darkgreen}{P(S(u)|u)=1}} \underbrace{P(m|u,i) P(Y|q,c,m)}_{\textcolor{darkgreen}{P(M(u,i)|u,i)=1}}} \label{eq:correlation_2}\\
&= \sum_{p\in \mathcal{P}}\sum_{c\in\mathcal{C}} P(p|i)\, P(c|p,S(u))\, P(Y|q,c,M(u,i)), \label{eq:correlation_3}
\end{align}
\end{subequations}
where we omit the superscript $t$ for simplicity. We derive Eq.(\textcolor{blue}{\ref{eq:correlation_1}}) by the law of total probability, and Eq.(\textcolor{blue}{\ref{eq:correlation_2}}) follows from Bayes' Theorem. Since $M(u,i)$ can take a value only if $U=u$ and $I=i$, we have $P(M(u,i)|u,i)=1$ and $P(S(u)|u)=1$, allowing us to remove the sum over $\mathcal{M}$ and $\mathcal{S}$. Therefore, we derive Eq.(\textcolor{blue}{\ref{eq:correlation_3}}).

\subsection{Deconfounded Training}
\label{sec:training}
To estimate the causal effect of all variables on the click action $Y$, we apply deconfounded training to remove the confounder structure in Figure~\textcolor{blue}{\ref{fig:causal_graph}(A)}, by the backdoor adjustment. 
Specifically, our training approach includes three key aspects: {\textbf{(i)}} We apply the backdoor adjustment to cut off the path $P\rightarrow I$ and estimate the causal effect of all variables. 
{\textbf{(ii)}} We model the gap between  {evolving personal popularity} and local popularity in the conformity effect to measure the influence of {evolving personal popularity}. 
{\textbf{(iii)}} We introduce the item quality proportional loss and overall objective function.

\vspace{3pt}
\noindent \textit{\textbf{Backdoor Adjustment.}}
After the description of the causal graph, we need to estimate the causal effect from local popularity to interactions $P\rightarrow Y$. 
Backdoor adjustment is commonly used to cut off the path $P\rightarrow I$, which eliminates the effect of local popularity through the path $P\rightarrow I\rightarrow M\rightarrow Y$~\cite{DBLP:conf/sigir/ZhangF0WSL021}. 
According to the backdoor adjustment theory~\cite{pearl2009causality}, we can apply the do-calculus $do(I=i)$ to remove the impact of $I$’s parent nodes (Figure~\textcolor{blue}{\ref{fig:causal_graph}(B)}). 
Backdoor adjustment on the basis of Eq.(\textcolor{blue}{\ref{eq:correlation_3}}) can be formulated as: 
\begin{subequations}
\begin{align}
&P(Y \mid U = u, do(I = i), Q = q) \notag\\
&= \sum_{p\in \mathcal{P}}\sum_{c\in\mathcal{C}} P(p|do(I=i)) P(c|S(u),p) P(Y|q,c,M(u,i)) \label{eq:do_calculus_1} \\
&= \sum_{p\in \mathcal{P}}\sum_{c\in\mathcal{C}} P(p) P(c|S(u),p) P(Y|q,c,M(u,i)) \label{eq:do_calculus_2} \\
&= \sum_{p\in\mathcal{P}} P(p) P(Y|q,C(S(u),p),M(u,i)) \label{eq:do_calculus_3}\\
&=\  \mathbb{E}_p[P(Y|q,C(S(u),p),M(u,i))] \label{eq:do_calculus_4}, 
\end{align}
\end{subequations}
where Eqs.(\textcolor{blue}{\ref{eq:do_calculus_1}},\textcolor{blue}{\ref{eq:do_calculus_2}}) follow the backdoor criterion~\cite{pearl2009causality}, as the only backdoor path, $I \leftarrow P \rightarrow C$, in the causal graph is blocked by $do(I=i)$.
Eqs.(\textcolor{blue}{\ref{eq:do_calculus_3}},\textcolor{blue}{\ref{eq:do_calculus_4}}) are derived because $C(S(u),p)$ can take a value only when $S=S(u)$ and $P=p$, indicating $P(C(S(u),p)|S(u),p)=1$, allowing us to remove the sum over $\mathcal{C}$. 

\vspace{3pt}
\noindent \textit{\textbf{{Evolving Personal Popularity}}}.
We design our model according to $P(Y|q,C(S(u),P),M(u,i))$ in Eq.(\textcolor{blue}{\ref{eq:do_calculus_4}}), as shown in the following: 
\begin{equation}
    \hat{y}_{ui}^t = \text{Tanh}(q_i+c_{ui}^t) \cdot \text{Softplus}(m_{ui}),
    \label{eq:prediction_score}
\end{equation}
where $q_i$ is a learnable parameter to measures the quality of item $i$, $c_{ui}$ is the conformity effect between user $u$ and item $i$, and $m_{ui}$ is the matching score that can be implemented by various recommendation models, such as MF~\cite{DBLP:conf/uai/RendleFGS09} and LightGCN~\cite{DBLP:conf/sigir/0001DWLZ020}. 
Tanh($\cdot$) and Softplus($\cdot$) are hyperbolic tangent and a softplus activation function, respectively, to make the model trained more stable.

The key of the conformity effect between user $u$ and the item $i$ lies in the design $C(S(u),P)$.
We define the conformity effect by the gap between  {evolving personal popularity}
and local popularity:
\begin{equation}
    c_{ui}^t = e^{-\alpha|s_u^t-p_i^t|} \cdot p_i^t \cdot \text{MLP}(i),
    \label{eq:conformity_effect}
\end{equation}
where $|s_u^t-p_i^t|$ is the consistency between  {evolving personal popularity} and local popularity. $e^{-\alpha|s_u^t-p_i^t|}$ is the {\textbf{consistency score}}, which quantifies the gap. When the gap is large, the consistency score is close to $0$. 
The parameter $\alpha$, called the {\textbf{consistency ratio}}, controls how the consistency score responds to the gap.



\vspace{3pt}
\noindent  \textit{\textbf{Objective Function}}.
Before giving the overall objective function, we should replace $\hat{y}^{t}_{ui}$ with $\mathbb{E}_p[\hat{y}^{t}_{ui}]$, as shown in Eq.(\textcolor{blue}{\ref{eq:do_calculus_4}}). However, since $\mathbb{E}_p[\hat{y}^{t}_{ui}]$ is difficult to compute because of the large space of local popularity $P$, we can approximate it as follows:
\begin{align}
    \mathbb{E}_p[\hat{y}^{t}_{ui}]
    &\approx \text{Tanh}(q_i+\mathbb{E}_p[c_{ui}^t]) \cdot \text{Softplus}(m_{ui}), \label{eq:approx_1} \\
    \mathbb{E}_p[c_{ui}^t] &\approx
    e^{-\alpha|s_u^t-\mathbb{E}_p[p_i^t]|} \cdot \mathbb{E}_p[p_i^t] \cdot \text{MLP}(i), \label{eq:approx_2}
\end{align}
where $\mathbb{E}_p[p_i^t]$ is the average local popularity in the train data, and the approximation of Eqs.(\textcolor{blue}{\ref{eq:approx_1}},\textcolor{blue}{\ref{eq:approx_2}}) are obtained from the Jensen’s inequality of non-linear function, as shown in Theorem~\textcolor{blue}{\ref{theorem:expectation_approximation}}.

\begin{theorem}[Jensen's inequality of non-linear function]
\label{theorem:expectation_approximation}
    If a random variable $X$ with the probability distribution $P(X)$ has the expectation $\mathbb{E}(X)$, given the non-linear function $f: \mathcal{G} \to \mathbb{R}$ where $\mathcal{G}$ is a closed subset of \,$\mathbb{R}$, following:
\begin{enumerate}
    \item $f$ is bounded on any compact subset of $\mathcal{G}$;
    \item $|f(x) - f(\mathbb{E}(X))| = O(|x - \mathbb{E}(X)|^\beta)$ at $x \to \mu$ for $\beta > 0$;
    \item $|f(x)| = O(|x|^\gamma)$ as $x \to +\infty$ for $\gamma \geq \beta$,
\end{enumerate}
then the inequality holds:
$
|\mathbb{E}[f(X)] - f(\mathbb{E}(X))| \leq T (\rho_\beta^\beta + \rho_\gamma^\gamma),
$
where 
$
\rho_\beta = \sqrt[\beta]{\mathbb{E}[|X - \mathbb{E}(X)|^\beta]},$ and $T$ does not depend on $P(X)$. The proof can be found in~\cite{gao2017bounds}.
\end{theorem}

To ensure the quality is proportional to global popularity, we propose a quality loss:
\begin{equation}
    \mathcal{L}_{Q} = \sum_{i\in \mathcal{I}}\sum_{j\sim p_n} -\text{log}(\sigma(\text{Sign}(d_i-d_j)({q}_{i}-{q}_{j}))),
    \label{eq:loss_quality}
\end{equation}
where $p_n$ is the negative sampling distribution, $d_i$ refers to the global popularity of item $i$. $\text{Sign}(\cdot)$ is the sign function, where $\text{Sign}(\cdot)=1$ if $d_i>d_j$, $\text{Sign}(\cdot)=-1$ if $d_i<d_j$, and $\text{Sign}(\cdot)=0$ if $d_i=d_j$. This sign function ensures that $q_i>q_j$ if $d_i>d_j$, and $q_i<q_j$ otherwise. 

Finally, we adopt the commonly used Bayesian Personalized Ranking (BPR) loss on the prediction score to train the model. 
\begin{align}
&\mathcal{L} = \mathcal{L}_{\text{BPR}} + \lambda \mathcal{L}_{Q}, \label{eq:over_loss} \\
&\mathcal{L}_{\text{BPR}} = \sum_{(u,i,t)\in \mathcal{D}}\sum_{j\sim p_n} -\text{log}(\sigma(\mathbb{E}_p[\hat{y}^{t}_{ui}]-\mathbb{E}_p[\hat{y}^{t}_{uj}])),
    \label{eq:loss_bpr}
\end{align}
where $\lambda$ is a hyperparameter to balance the quality loss and BPR loss, $\mathbb{E}_p[\hat{y}^{t}_{ui}]$ is the approximated expectation of $\hat{y}^{t}$ in Eq.(\textcolor{blue}{\ref{eq:approx_1}}), and $p_n$ is the negative sampling distribution. 

\subsection{Intervened Inference}
\label{sec:inference}

After deconfounded training, our model captures the causal effect of each variable on the click action $Y$. Based on this, we can apply the intervention (by do-calculus) to local popularity and  {evolving personal popularity} to achieve popularity debiasing. Specifically, owing to $P(Y|U,do(I=i),Q)$ and the approximation of $\mathbb{E}_p[\hat{y}^{t}_{ui}]$, we can estimate the causal effect of local popularity. 
We can adjust the causal effect of local popularity by intervention and set $P = {p}^*$ during the inference stage, as shown in Figure~\textcolor{blue}{\ref{fig:causal_graph}(C)}, formulated as: 
\begin{equation}
    P(Y|U,do(I=i),Q,do(P={p}^*))=P(Y|U,do(I=i),Q,{p}_i^*),
    \label{eq:intervene}
\end{equation}
where ${p}_i^*$ denotes the intervened popularity value for item $i$. 

However, finding a suitable adjustment ${p}^*$ is challenging. It has been shown that completely eliminating popularity bias by setting ${p}^*$ as the average value $\mathbb{E}_p[p_i^t]=\sum_{(\cdot,i,t)\in \mathcal{D}}({p_i^t}/{|\mathcal{D}|})$ degrades performance~\cite{DBLP:journals/tkde/ZhaoCZHCZW23}.
Other research~\cite{DBLP:journals/tkde/ZhuZFYWH24} uses grid search for tuning $p^*$, which is difficult due to the infinite search space and disregard of the nature of popularity evolution, which consequently inaccurately captures popularity bias in testing time. 
We assume that temporal evolution could guide us in finding a suitable intervention. 
We apply a simple time series forecasting method to forecast future local popularity. Specifically, we adjust popularity based on the gradient (direction) of moving average lines:
\begin{equation}
    {p}_i^* = p_i^T + \frac{\partial \text{MA}}{\partial t}(p_i^T) \cdot \Delta_i^T,
    \label{eq:popularity_drift}
\end{equation}
where $\frac{\partial \text{MA}}{\partial t}(\cdot)$ represents the gradient of the moving average line at a given point, which indicates the direction of future evolution. The $\Delta_i^T$ is the time step inferred from the last time stage $T$.

\begin{table*}[h]
\centering
\scriptsize
\caption{Performance comparison with LightGCN as a backbone.
\textbf{\textcolor{brightmaroon}{Bold}} and \underline{\textcolor{violet}{underline}} emphasize the best and second-best results, respectively. 
$\dag$ indicates a statistically significant improvement over the best-performing 
baseline using t-test with $p<0.05$. 
}
\vspace{-8pt}
\label{tab:performance_comparison_lightgcn}
\resizebox{1\textwidth}{!}
{
{
\begin{tabular}{l|ccc|ccc|ccc|c}
\toprule
Datasets    & \multicolumn{3}{c|}{Ciao}       & \multicolumn{3}{c|}{ Amazon-Music} &  \multicolumn{3}{c|}{Douban-movie}  & Avg.   \\ 
Metrics     & Rec@20 & Pre@20 & NDCG@20  & Rec@20 & Pre@20 & NDCG@20  &  Rec@20 & Pre@20 & NDCG@20  & Rank  \\ 
 \midrule \midrule
LightGCN~\cite{DBLP:conf/sigir/0001DWLZ020}     & 0.0181 & 0.0083 & 0.0091 & 0.0528 & 0.0086 & 0.0090 & 0.0227 & 0.0363 & 0.0390 & 8.83 \\
IPS~\cite{DBLP:conf/ijcai/JoachimsSS18} & 0.0249 & 0.0076 & 0.0091 & 0.0481 & 0.0074 & 0.0076 & 0.0240 & 0.0376 & 0.0385 & 9.39 \\
DICE~\cite{DBLP:conf/www/ZhengGLHLJ21}         & \underline{\textcolor{violet}{0.0255}} & 0.0090 & 0.0083 & 0.0514 & 0.0077 & 0.0086 & 0.0232 & 0.0375 & 0.0389 & 8.33 \\
MACR~\cite{DBLP:conf/kdd/WeiFCWYH21}         & 0.0220 & 0.0063 & 0.0074 & 0.0522 & 0.0083 & 0.0098 & 0.0315 & 0.0472 & 0.0497 & 8.67 \\
PD~\cite{DBLP:conf/sigir/ZhangF0WSL021}           & 0.0241 & 0.0069 & 0.0069 & 0.0495 & 0.0070 & 0.0080 & 0.0332 & 0.0481 & 0.0530 & 8.94 \\
PDA~\cite{DBLP:conf/sigir/ZhangF0WSL021}          & 0.0204 & 0.0068 & 0.0068 & 0.0698 & 0.0115 & 0.0126 & 0.0409 & 0.0559 & 0.0624 & 7.67 \\
UDIPS~\cite{DBLP:conf/sigir/LuoW23}        & 0.0241 & 0.0086 & 0.0087 & 0.0503 & 0.0075 & 0.0078 & 0.0259 & 0.0381 & 0.0389 & 8.78 \\
PARE~\cite{DBLP:conf/cikm/JingZ0023}         & 0.0227 & 0.0069 & 0.0069 & 0.0697 & 0.0121 & 0.0130 & 0.0426 & 0.0574 & 0.0643 & 6.67 \\
TIDE~\cite{DBLP:journals/tkde/ZhaoCZHCZW23}         & 0.0234 & \underline{\textcolor{violet}{0.0091}} & \underline{\textcolor{violet}{0.0102}} & 0.0702 & 0.0129 & 0.0140 & 0.0482 & \textbf{\textcolor{brightmaroon}{0.0648}} & \underline{\textcolor{violet}{0.0726}} & 3.89 \\ 
PPAC~\cite{DBLP:conf/www/NingC0KHH024}         & 0.0245 & 0.0090 & 0.0097 & \underline{\textcolor{violet}{0.0705}} & \underline{\textcolor{violet}{0.0130}} & \underline{\textcolor{violet}{0.0142}} 
& \underline{\textcolor{violet}{0.0485}} & 0.0637 & 0.0722 & \underline{\textcolor{violet}{3.72}} \\ 
 \midrule \midrule
CausalEPP       & \textbf{\textcolor{brightmaroon}{0.0293}}$^\dag$ & \textbf{\textcolor{brightmaroon}{0.0097}}$^\dag$ & \textbf{\textcolor{brightmaroon}{0.0112}}$^\dag$ & \textbf{\textcolor{brightmaroon}{0.0723}}$^\dag$ & \textbf{\textcolor{brightmaroon}{0.0135}}$^\dag$ & \textbf{\textcolor{brightmaroon}{0.0145}}$^\dag$ & \textbf{\textcolor{brightmaroon}{0.0512}}$^\dag$ & \underline{\textcolor{violet}{0.0646}} & \textbf{\textcolor{brightmaroon}{0.0736}}$^\dag$ & \textbf{\textcolor{brightmaroon}{2.11}} \\ 
Impv.       &  14.9\% & 6.6\% & 9.8\% & 2.6\% & 3.9\% & 2.1\% & 5.6\% & -0.3\% & 1.4\% & -\\
\midrule
CausalEPP-Inv        & \textbf{\textcolor{brightmaroon}{0.0303}}$^\dag$  & \textbf{\textcolor{brightmaroon}{0.0108}}$^\dag$  & \textbf{\textcolor{brightmaroon}{0.0117}}$^\dag$  & \textbf{\textcolor{brightmaroon}{0.0741}}$^\dag$  & \textbf{\textcolor{brightmaroon}{0.0142}}$^\dag$  & \textbf{\textcolor{brightmaroon}{0.0149}}$^\dag$  & \textbf{\textcolor{brightmaroon}{0.0514}}$^\dag$  & \textbf{\textcolor{brightmaroon}{0.0649}}$^\dag$  & \textbf{\textcolor{brightmaroon}{0.0740}}$^\dag$  & \textbf{\textcolor{brightmaroon}{1.00}} \\
Impv.       &  18.8\% & 18.7\% & 14.7\% & 5.1\% & 9.2\% & 4.9\% & 6.0\% & 0.2\% & 1.9\% & - \\
\bottomrule
\end{tabular}
}}
\end{table*}

Similarly, the {personal popularity} also evolves over time. 
To better adjust the conformity effect during the inference stage, we also forecast the future  {personal popularity}:
\begin{equation}
    {s}_u^* = s_u^T + \frac{\partial \text{MA}}{\partial t}(s_u^T) \cdot \Delta_u^T,
    \label{eq:sensitivity_drift}
\end{equation}
where $\Delta_u^T$ and $\Delta_i^T$ may have different values, since they may have different time series trends. 
It is important to consider both the temporal evolution of  {personal popularity} and local popularity because ignoring one of them can lead to inaccurate recommendation. 
Therefore, the overall inference stage is described by $\hat{y}_{ui}^*$ and the intervened conformity effect $c_{ui}^*$ is given by: 
\begin{align}
    &\hat{y}_{ui}^* = \text{Tanh}(q_i+{c}_{ui}^*)\cdot \text{Softplus}(m_{ui}), \\
    &c_{ui}^* = e^{-\alpha|s_u^*-p_i^*|} \cdot p_i^* \cdot \text{MLP}(i). \label{eq:intervention}
\end{align}
From this formulation, we intervene the conformity effect indirectly by intervening in local popularity and  {evolving personal popularity}. 

\IncMargin{1.0em}
\begin{algorithm}[h]
\small
\caption{CausalEPP}
\label{alg:link_prediction_algorithm}

\SetKwData{Left}{left}\SetKwData{This}{this}\SetKwData{Up}{up}
\SetKwFunction{Union}{Union}
\SetKwInOut{Input}{Input}

\Indm
\Input{dataset $\mathcal{D}=\{(u, i, t)|u\in\mathcal{U}, i\in\mathcal{I}\}$; loss balance weight $\lambda$; sliding window size $w_1$; intervention parameters $\alpha$, $\Delta_u^T$, $\Delta_i^T$, $w_2$.}
\BlankLine
\Indp

Calculating the average local popularity $\mathbb{E}_p[p_i^t]=\sum_{(\cdot,i,t)\in \mathcal{D}}\frac{p_i^t}{|\mathcal{D}|}$. 

\tcp{\textcolor{darkgreen}{\textbf{Deconfounded Training}}}
\For{$T$ Epochs}{
\For{$(u, i, t) \in \mathcal{D}$}{
Get personal popularity $s_u^t$ from Eq.(\ref{eq:popularity_sensitivity}).

Approximate the conformity effect $\mathbb{E}_p[c_{ui}^t]$ by Eq.(\ref{eq:approx_2}).

Approximate the click score $\mathbb{E}_p[\hat{y}_{ui}^t]$ by Eq.(\ref{eq:approx_1}). 

Calculate the overall loss $\mathcal{L}=\mathcal{L}_{\text{BPR}}+\lambda\mathcal{L}_{Q}$.


Update the model by gradient descent. 
}
}

\tcp{\textcolor{darkgreen}{\textbf{Temporal Evolution based Inference}}}
Get local popularity $p_i^T=\frac{|\mathcal{D}_i^{T-w_1,T}|}{|\mathcal{D}^{T-w_1,T}|}$.

Get personal popularity $s_u^T = \frac{|\{(u,i):i\in\mathcal{D}_u^{T-w_1,T} \ \text{and}\  p_i^T>\hat{p}^T\}|}{|\mathcal{D}_u^{T-w_1,T}|}$.

Calculating gradient of moving average line for the local popularity $\frac{\partial \text{MA}}{\partial t}(p_i^T)$ and personal popularity $\frac{\partial \text{MA}}{\partial t}(s_u^T)$ by Eq.(\ref{eq:moving_average}).

Intervene local popularity $p_i^*=p_i^T + \frac{\partial \text{MA}}{\partial t}(p_i^T) \cdot \Delta_i^T$.

Intervene popularity sensitivity $s_u^*=s_u^T + \frac{\partial \text{MA}}{\partial t}(s_u^T) \cdot \Delta_u^T$.

Recommend items by set $P=p_i^*$ and $S=s_u^*$.

\end{algorithm}
\DecMargin{1.0em}

\section{Experiment}

To comprehensively evaluate the effectiveness of CausalEPP, we conduct extensive experiments
to address the following questions.

\begin{itemize}[leftmargin=12pt]
    \item \textbf{RQ1.} 
    How does the performance of our CausalEPP compare to various state-of-the-art baseline methods?
    \item \textbf{RQ2.} 
    What are the effect of (1) the proposed evolving personal popularity, (2) temporal evolution-guided inference, and (3) the learned item quality?
    \item \textbf{RQ3.} How do hyperparameters affect CausalEPP?
    \item \textbf{RQ4.} Does CausalEPP successfully mitigate popularity bias?
\end{itemize}

\subsection{Experimental Settings}
\label{subsec:experiment_setting}

\noindent \textit{\textbf{Evaluation Datasets.}} Following previous works~\cite{DBLP:conf/sigir/ZhangF0WSL021,DBLP:journals/tkde/ZhaoCZHCZW23}, we chose three well-known datasets,  Ciao~\cite{DBLP:conf/icwsm/LakkarajuML13}, Amazon-Music~\cite{DBLP:conf/icwsm/LakkarajuML13}, and Douban-Movie~\cite{DBLP:conf/wsdm/Song0WCZT19}.
The datasets contain user rating records in chronological order, and we conduct 5-core filtering for Ciao and Amazon-Music, and 10-core filtering for Douban-Movie.
We chronologically split each dataset into 10 parts.
The first nine parts are used for training, while the last part is evenly divided for validation and testing.
We divide time into $100$ time steps and compute local popularity and evolving personal popularity for each step $t$ over the interval $[t-w_1, t]$. 
The statistics are provided in Table~\textcolor{blue}{\ref{tab:dataset}}.

\begin{table}[h]
\centering
\scriptsize
\caption{Experimental data statistics.}
\vspace{-8pt}
\label{tab:dataset}
\resizebox{0.47 \textwidth}{!}
{
\setlength{\tabcolsep}{1.2mm}
{
\begin{tabular}{l|ccccc}
\toprule
 Datasets            & \# User  & \# Item    & \# Interaction      & Date  & Step time  \\ \midrule \midrule
Ciao            & 5,868 & 10,724 & 143,217 & 2000.5-2011.4 & 37 days \\
Amazon-Music      & 5,541 & 3,568 & 64,706 & 1998.04-2014.7 & 57 days \\
Douban-Movie    & 48,799 & 26,813  & 7,409,868 & 2010.1-2017.3 & 24 days \\
\bottomrule
\end{tabular}
}}
\vspace{-8pt}
\end{table}

\vspace{3pt}
\noindent   \textit{\textbf{Baselines.}} For a comprehensive comparison, we adopt ten baselines, including 
\textbf{{(i)}} Inverse Propensity Score-based: IPS~\cite{DBLP:conf/ijcai/JoachimsSS18,DBLP:conf/icml/SchnabelSSCJ16} and UDIPS~\cite{DBLP:conf/sigir/LuoW23};  
\textbf{{(ii)}} Counterfactual inference-based: MACR~\cite{DBLP:conf/kdd/WeiFCWYH21} and PPAC~\cite{DBLP:conf/www/NingC0KHH024}; 
\textbf{{(iii)}} Deconfounded-based: DICE~\cite{DBLP:conf/www/ZhengGLHLJ21}, PD~\cite{DBLP:conf/sigir/ZhangF0WSL021}, and TIDE~\cite{DBLP:journals/tkde/ZhaoCZHCZW23}; 
\textbf{{(iv)}} Distribution evolution-based: PARE~\cite{DBLP:conf/cikm/JingZ0023} and PDA~\cite{DBLP:conf/sigir/ZhangF0WSL021}. 

\vspace{3pt}
\noindent  \textit{\textbf{Evaluation Metrics.}} We evaluate the capability of models to predict future user clicks by grading items with which users have not interacted and testing whether the top-K items are clicked, using Recall@K (Rec@K), Precision@K (Pre@K), and NDCG@K~\cite{DBLP:journals/tois/JarvelinK02}.

\vspace{3pt}
\noindent  \textit{\textbf{Implementation Details.}} We adopt MF~\cite{DBLP:conf/uai/RendleFGS09} and LightGCN~\cite{DBLP:conf/sigir/0001DWLZ020} as our backbones. 
We implement all models in PyTorch and optimize them using the Adam optimizer with a batch size of 8,192.
Hyperparameters are tuned through grid search based on the validation data.
We set hyperparameters as follows: $\alpha=0.5$, $\lambda=0.2$, $w_1=6$ months, quantile for high-popularity items $=20\%$, $w_2=10\%$, $\Delta_i^T=5$, and $\Delta_u^T=10$. 
Experiments are conducted on a single machine with an Intel i9-10850K processor and Nvidia RTX 3090Ti (24GB) GPUs.

\begin{table*}[h]
\centering
\scriptsize
\caption{Performance comparison with MF as a backbone.
The notations are the same as those in Table~\textcolor{blue}{\ref{tab:performance_comparison_lightgcn}}.
}
\vspace{-8pt}
\label{tab:performance_comparison_mf}
\resizebox{1.0\textwidth}{!}
{
{
\begin{tabular}{l|ccc|ccc|ccc|c}
\toprule
 Datasets            & \multicolumn{3}{c|}{Ciao}       & \multicolumn{3}{c|}{ Amazon-Music} &  \multicolumn{3}{c}{Douban-movie}   & Avg.  \\ 
 Metrics            & Rec@20 & Pre@20 & NDCG@20   & Rec@20 & Pre@20 & Ndcg@20 &  Rec@20 & Pre@20 & Ndcg@20    & Rank \\ 
 \midrule \midrule
MF~\cite{DBLP:conf/uai/RendleFGS09}           & 0.0107 & 0.0076 & 0.0086 & 0.0362 & 0.0068 & 0.0080 & 0.0223 & 0.0342  &  0.0370  & 10.44 \\
IPS~\cite{DBLP:conf/ijcai/JoachimsSS18}       & 0.0109 & 0.0068 & 0.0078 & 0.0378 & 0.0063 & 0.0071 &  0.0220  & 0.0337 &  0.0366 & 11.11 \\
DICE~\cite{DBLP:conf/www/ZhengGLHLJ21}         & 0.0145 & 0.0103 & 0.0110 & 0.0357 & 0.0058 & 0.0060 &  0.0202 &  0.0323 & 0.0343 & 11.00 \\
MACR~\cite{DBLP:conf/kdd/WeiFCWYH21}         & 0.0182 & 0.0115 & 0.0123 & 0.0463  & 0.0082 & 0.0096 &  0.0388  & 0.0526  &  0.0541 & 7.00 \\
PD~\cite{DBLP:conf/sigir/ZhangF0WSL021}           & 0.0177 & 0.0110 & 0.0112 & 0.0418 & 0.0070 & 0.0076 &  0.0355  & 0.0465 &  0.0520 & 8.33 \\
PDA~\cite{DBLP:conf/sigir/ZhangF0WSL021}          & 0.0189 & 0.0144 & \textbf{\textcolor{brightmaroon}{0.0159}} & 0.0656  & 0.0111 & 0.0119 &  0.0408  & 0.0534  &  0.0596 & 5.22 \\
UDIPS~\cite{DBLP:conf/sigir/LuoW23}        & 0.0111 & 0.0078 & 0.0097 & 0.0405 & 0.0072 & 0.0088 & 0.0247 & 0.0381 & 0.0372 & 9.11 \\
PARE~\cite{DBLP:conf/cikm/JingZ0023}         & 0.0194 & 0.0142 & 0.0147 & 0.0662 & 0.0123 & 0.0126 & 0.0420 & 0.0551 & 0.0614 & 5.22 \\
TIDE~\cite{DBLP:journals/tkde/ZhaoCZHCZW23}         & 0.0244 & \underline{\textcolor{violet}{0.0148}} & {0.0154}
& {0.0837} & \underline{\textcolor{violet}{0.0148}} & \underline{\textcolor{violet}{0.0164}}
& \underline{\textcolor{violet}{0.0473}} & \underline{\textcolor{violet}{0.0556}} & \underline{\textcolor{violet}{0.0631}} & \underline{\textcolor{violet}{3.17}} \\ 
PPAC~\cite{DBLP:conf/www/NingC0KHH024}         & \underline{\textcolor{violet}{0.0260}} & \textbf{\textcolor{brightmaroon}{0.0149}} & \underline{\textcolor{violet}{0.0156}} & \underline{\textcolor{violet}{0.0843}} & 0.0142 & 0.0158 & 0.0465 & 0.0553 & 0.0622 & 3.22 \\ 
 \midrule \midrule
CausalEPP     & \textbf{\textcolor{brightmaroon}{0.0313}}$^\dag$ & 0.0143 & 0.0150 & \textbf{\textcolor{brightmaroon}{0.0857}}$^\dag$ & \textbf{\textcolor{brightmaroon}{0.0153}}$^\dag$ & \textbf{\textcolor{brightmaroon}{0.0169}}$^\dag$ & \textbf{\textcolor{brightmaroon}{0.0485}}$^\dag$ & \textbf{\textcolor{brightmaroon}{0.0568}}$^\dag$ & \textbf{\textcolor{brightmaroon}{0.0642}}$^\dag$ & \textbf{\textcolor{brightmaroon}{2.67}}\\ 
Impv.       &  20.4\% & -4.0\% & -5.6\% & 1.7\% & 3.3\% & 3.0\% & 2.5\% & 2.2\% & 1.7\% & - \\
\midrule
CausalEPP-Inv        & \textbf{\textcolor{brightmaroon}{0.0322}}$^\dag$ & {0.0147} & {0.0154} & \textbf{\textcolor{brightmaroon}{0.0866}}$^\dag$ & \textbf{\textcolor{brightmaroon}{0.0156}}$^\dag$ & \textbf{\textcolor{brightmaroon}{0.0170}}$^\dag$ & \textbf{\textcolor{brightmaroon}{0.0490}}$^\dag$ & \textbf{\textcolor{brightmaroon}{0.0576}}$^\dag$ & \textbf{\textcolor{brightmaroon}{0.0664}}$^\dag$ & \textbf{\textcolor{brightmaroon}{1.50}} \\
Impv.       &  23.8\% & -1.3\% & -3.1\% & 2.7\% & 5.4\% & 3.7\% & 3.6\% & 3.6\% & 5.2\% & - \\
\bottomrule
\end{tabular}
}}
\end{table*}

\subsection{Performance Comparison (RQ1)}
We present the empirical performance comparison of baselines in Tables~\textcolor{blue}{\ref{tab:performance_comparison_lightgcn}} and \textcolor{blue}{\ref{tab:performance_comparison_mf}} with LightGCN and MF backbones, respectively. 
In these experiments, CausalEPP operates without intervention during the inference stage, which means that 
it does not perform intervention during inference.
CausalEPP achieves outstanding performance (average ranks of $2.11$ and $2.67$) across all metrics for the Ciao, Amazon Music, and Douban movie datasets. 
With the LightGCN backbone, CausalEPP improves performance by 6.67\% over TIDE and 6.70\% over PPAC. 
With the MF backbone, CausalEPP improves performance by an average of 4.17\% over TIDE and 4.36\% over PPAC. 
These results show that CausalEPP, by separating item quality from popularity and considering the consistency between {evolving personal popularity} and local popularity, effectively captures users' true interests and achieve better recommendation. 

We also show the performance of CausalEPP-Inv, which applies the intervention to CausalEPP during inference. 
CausalEPP-Inv adjusts the conformity effect by Eq.(\textcolor{blue}{\ref{eq:intervention}}), which accounts for the evolution of {personal popularity} and local popularity. 
By forecasting the future value of two evolutions, CausalEPP-Inv adaptively modifies the conformity effect for various users and items, improving the precision of the recommendation. 
The results show that the evolution-guided intervention consistently improves the performance, $3.45\%$ for the LightGCN and $1.98\%$ for the MF. 
This indicates that the proposed invention improves the results of the recommendation by forecasting the consistency score.

\subsection{Ablation Study (RQ2)}
In this section, we list ablation studies on the LightGCN backbone and discuss the effectiveness of three main components, including  {evolving personal popularity}, quality, and temporal evolution, as shown in Table~\textcolor{blue}{\ref{tab:ablation_studies_1}}. 
Since  {evolving personal popularity} affects both the training and inference stages, we present ablation results in both stages.
In contrast, quality affects only the training stage, 
while temporal evolution affects only the inference stage.

\vspace{3pt}
\noindent  \textit{\textbf{(1) Effect of  {Evolving Personal Popularity}.}}
The  {evolving personal popularity} can work only through the consistency score in Eq.(\textcolor{blue}{\ref{eq:conformity_effect}}). We can analyze its impact by removing the consistency score, denoted as ``\textbf{w/o consistency}''. 
For CausalEPP, ``\textbf{w/o consistency}'' indicates that removing the consistency score (the path $S\rightarrow C$) during training, makes the conformity effect independent of  {evolving personal popularity}. 
The results 
show a performance drop of $-3.18\%$. 
For CausalEPP-Inv, ``\textbf{w/o consistency}'' indicates that no intervention is performed on personal popularity during inference. 
The results 
show a performance drop of $-1.58\%$. 
This outcome demonstrates that adjusting the conformity effect based on {evolving personal popularity}, or the consistency score, captures a deeper insight into popularity bias, leading to a more accurate prediction.

\begin{figure}[ht]
	\centering
        \includegraphics[scale=0.35]{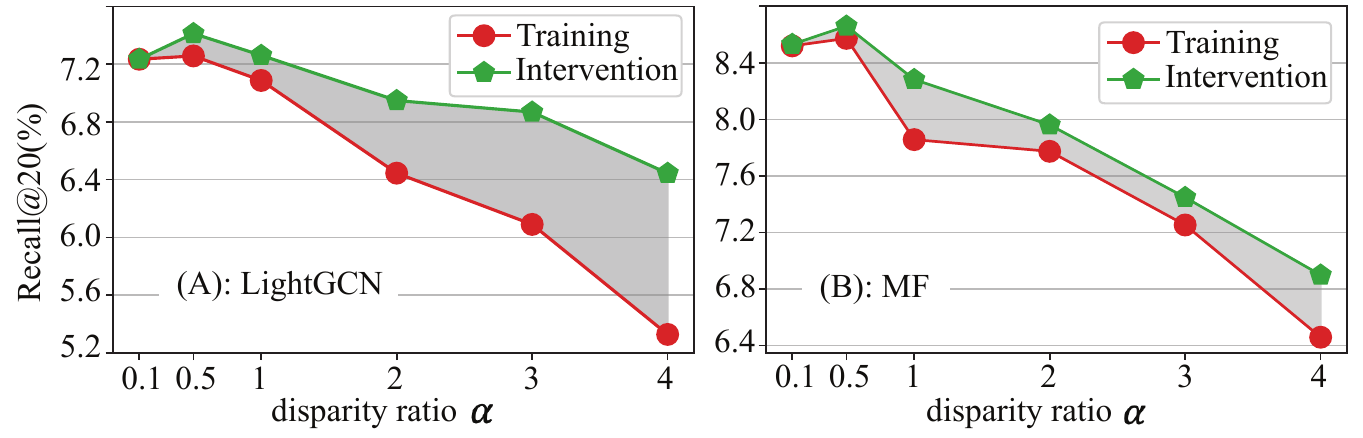}
        \vspace{-8pt}
	\caption{We illustrate the performance (\%) on the Amazon-Music dataset for both LightGCN and MF backbones with varying consistency ratios $\alpha$.
    The shaded area highlights the performance gain brought from the intervention.}
    \label{fig:param_alpha_amazonmusic}
\end{figure}

We further investigate the influence of the parameter $\alpha$ in Eq.(\textcolor{blue}{\ref{eq:conformity_effect}}). 
In Figure~\textcolor{blue}{\ref{fig:param_alpha_amazonmusic}}, we evaluate the performance change during the training (red line) and inference (green line) stages with various $\alpha$. 
We observe that the intervention consistently improves the trained model's performance. However, as $\alpha$ increases, the performance declines because the consistency score becomes overly sensitive to the gap $|s_u - p_i|$.
Furthermore, different values of $\alpha$ lead to varying degrees of improvement from the intervention, verifying the effect of the consistency score and  {evolving personal popularity}.

\begin{table*}[ht]
\centering
\scriptsize
\caption{Ablation study during training and inference with LightGCN as a backbone, where CausalEPP means during training and CausalEPP-Inv during inference. 
\textbf{\textcolor{darkgreen}{Bold}} denotes the improvements from the original model CausalEPP. 
}
\vspace{-8pt}
\label{tab:ablation_studies_1}
\resizebox{1\textwidth}{!}
{
{
\begin{tabular}{l|cc|cc|cc}
\toprule
\multicolumn{1}{c|}{Datasets}   & \multicolumn{2}{c|}{Ciao}       & \multicolumn{2}{c|}{ Amazon-Music} &  \multicolumn{2}{c}{Douban-movie} \\ 
\multicolumn{1}{c|}{Metrics}    & Rec@20 & NDCG@20  & Rec@20 & NDCG@20  &  Rec@20 & NDCG@20  \\ 
 \midrule \midrule
{CausalEPP}   & {0.0293} & {0.0112} & {0.0723} & {0.0145} & {0.0512} &{0.0736} \\
-w/o quality   & {0.0239} \textbf{\textcolor{darkgreen}{(-18.43\%)}} & {0.0100} \textbf{\textcolor{darkgreen}{(-10.71\%)}} & {0.0696} \textbf{\textcolor{darkgreen}{(-3.73\%)}}& {0.0130} \textbf{\textcolor{darkgreen}{(-10.34\%)}} & {0.0477} \textbf{\textcolor{darkgreen}{(-6.83\%)}} &{0.0714} \textbf{\textcolor{darkgreen}{(-2.98\%)}} \\
-w/o consistency   & {0.0277} \textbf{\textcolor{darkgreen}{(-5.46\%)}} & {0.0104} \textbf{\textcolor{darkgreen}{(-7.14\%)}} & {0.0710} \textbf{\textcolor{darkgreen}{(-1.80\%)}} & {0.0142} \textbf{\textcolor{darkgreen}{(-2.07\%)}} & {0.0504} \textbf{\textcolor{darkgreen}{(-1.56\%)}} &{0.0728} \textbf{\textcolor{darkgreen}{(-1.09\%)}} \\
\midrule
\midrule
{CausalEPP-Inv} & 0.0303  & 0.0117  & 0.0741  & 0.0149  & 0.0514  & 0.0740 \\
-eliminate p & 0.0144 \textbf{\textcolor{darkgreen}{(-50.85\%)}} & 0.0083 \textbf{\textcolor{darkgreen}{(-25.89\%)}} & 0.0514 \textbf{\textcolor{darkgreen}{(-28.90\%)}} & 0.0084 \textbf{\textcolor{darkgreen}{(-42.06\%)}} & 0.0253 \textbf{\textcolor{darkgreen}{(-50.58\%)}} & 0.0377 \textbf{\textcolor{darkgreen}{(-48.77\%)}} \\
-w/o evolution & 0.0285 \textbf{\textcolor{darkgreen}{(-5.94\%)}} & 0.0111 \textbf{\textcolor{darkgreen}{(-5.13\%)}} & 0.0718 \textbf{\textcolor{darkgreen}{(-3.10\%)}} & 0.0141 \textbf{\textcolor{darkgreen}{(-5.37\%)}} & 0.0499 \textbf{\textcolor{darkgreen}{(-2.92\%)}} & 0.0725 \textbf{\textcolor{darkgreen}{(-2.03\%)}} \\
-w/o consistency & 0.0296 \textbf{\textcolor{darkgreen}{(-2.31\%)}} &  0.0114 \textbf{\textcolor{darkgreen}{(-2.56\%)}}  & 0.0726 \textbf{\textcolor{darkgreen}{(-2.02\%)}} & 0.0146 \textbf{\textcolor{darkgreen}{(-2.01\%)}} & 0.0513 \textbf{\textcolor{darkgreen}{(-0.19\%)}} & 0.0737 \textbf{\textcolor{darkgreen}{(-0.41\%)}} \\
\bottomrule
\end{tabular}
}}
\end{table*}

\vspace{3pt}
\noindent  \textit{\textbf{(2) Effect of Temporal Evolution.}}
To evaluate the effect of evolution, we compare different intervention strategies to CausalEPP-Inv during inference stage: ``\textbf{eliminate p}'' indicates totally eliminating the popularity bias (setting $p^*$ to $\mathbb{E}[p]$) and ``\textbf{w/o evolution}'' replaces temporal evolution to regular intervention (grid searching $p^*$ and $s^*$ directly from $\{0.1, 0.2, ..., 1, 5, 10, ..., 50\}$). 
The results of ``\textbf{eliminate p}'' with a performance drop of $-41.18\%$ suggest that simply eliminating popularity bias negatively impacts the recommendation. 
``\textbf{w/o evolution}'' with no temporal information considered presents a performance drop of $-4.08\%$, showing that the temporal evolution intervention more accurately captures the user preference since regular intervention fails to utilize the tendency of evolution.

\vspace{3pt}
\noindent  \textit{\textbf{(3) Effect of Item Quality.}}
To evaluate whether the causal effect of quality is captured, we remove the quality disentangling component (path $Q\rightarrow Y$), referred to as ``\textbf{w/o quality}'' in Table~\textcolor{blue}{\ref{tab:ablation_studies_1}}. 
The results of ``\textbf{w/o quality}'' show a performance drop of $-8.84\%$, supporting the hypothesis that disentangling quality can improve recommendation accuracy. 
This result indicates that removing the quality component causes the model to lose its ability to identify the benign aspects of popularity, resulting in the conformity effect becoming unreliable as it entangles both the benign and harmful parts of popularity.

\begin{figure}[h]
	\centering
	\includegraphics[scale=0.36]{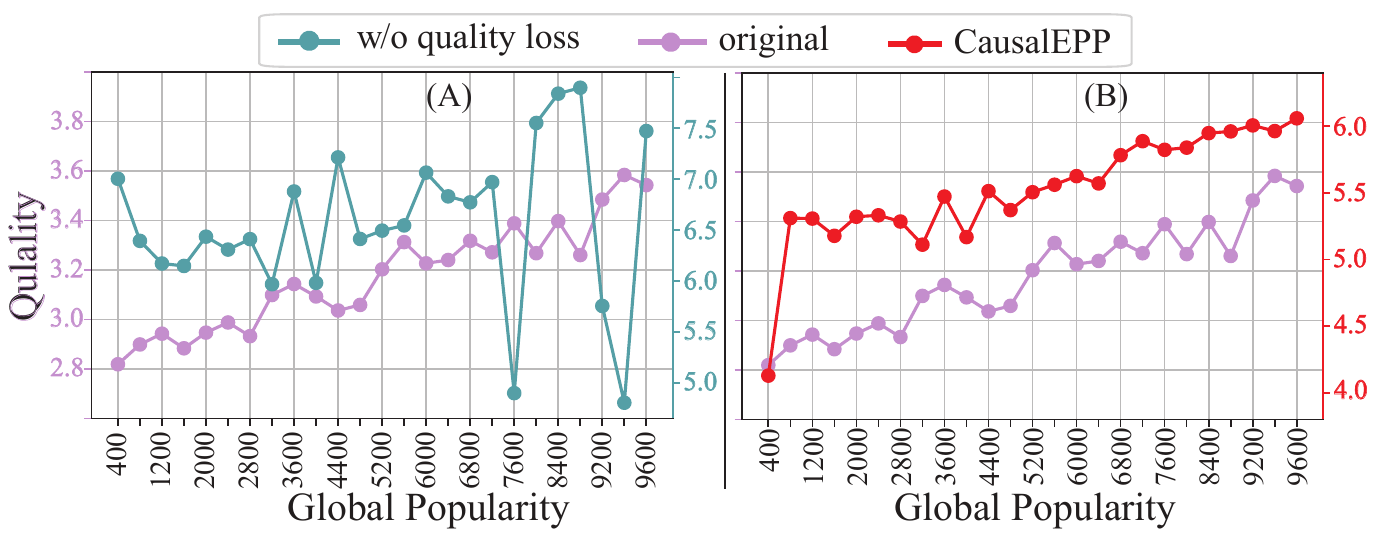}
    \vspace{-20pt}
	\caption{The ablation analysis of quality across global popularity on the Douban-Movie dataset. ``w/o quality loss'' indicates the quality loss was removed. ``original'' indicates the ground truth quality (the average rating of each group of items) and ``CausalEPP'' indicates the quality of CausalEPP. }
\label{fig:quality_analysis}
\end{figure}

\begin{figure*}[ht]
	\centering
        \includegraphics[scale=0.33]{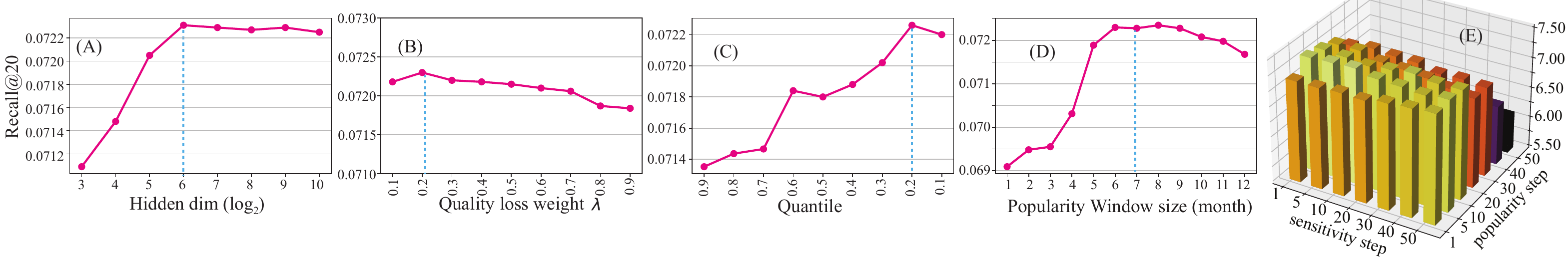}
        \vspace{-8pt}
	\caption{Parameter sensitive performance of recall@20 on the Amazon-music dataset with LightGCN as a backbone. (A): varying quantile, which defines the high-popularity items. (B): the hidden dimension of LightGCN.}
    \label{fig:param_sens}
\end{figure*}

Furthermore, to better understand the quality component, we evaluate the effectiveness of the quality loss. 
In Figure~\textcolor{blue}{\ref{fig:quality_analysis}(A)}, we compare the learned quality without the quality loss, and in Figure~\textcolor{blue}{\ref{fig:quality_analysis}(B)}, we compare the learned quality with it. 
From these figures, we observe that simply disentangling quality cannot capture a meaningful quality value, and our quality loss could capture the trend of original quality. 
These findings support the notion that our quality loss helps to learn meaningful quality.

\subsection{Hyper-parameter Study (RQ3)}
\label{sec:param_sens}

\noindent  \textit{\textbf{Training Stage.}}
In Figure~\textcolor{blue}{\ref{fig:param_sens}(A-D)}, we provide hyper-parameter sensitivity experiments for ``hidden dim'', 
``loss balance weight $\lambda$'', 
``high-popularity quantile in Eq.(\textcolor{blue}{\ref{eq:popularity_sensitivity}})'', 
and ``popularity window size $w_1$'' on Amazon-music. 
We observe that dimension 64 is enough to capture user click behavior. 
For $\lambda$, it is relatively insensitive to our selection, so we set it to $0.2$. 
The choice of quantile affects the definition of high-popularity items, which in turn influences evolving personal popularity, so we set the quantile to 20\%.
The ``$w_1$'' refers to the size of the sliding window for local popularity in Eq.(\textcolor{blue}{\ref{eq:popularity_bias}}). It directly affects local popularity and indirectly influences evolving personal popularity through $\hat{p}_i^t$, so we set it to $6$ months.



\vspace{3pt}
\noindent  \textit{\textbf{Inference Stage.}} 
In Figure~\textcolor{blue}{\ref{fig:param_sens}(E)}, we tune ``local popularity step $\Delta_i^T$'' and ``{evolving personal popularity} step $\Delta_u^T$''. 
Both the gradient and the step $\Delta^T$ determine the future value of evolution in Eqs.(\textcolor{blue}{\ref{eq:popularity_drift}}) and (\textcolor{blue}{\ref{eq:sensitivity_drift}}). 
From the results, the optimal time steps are set to $\Delta_i^T=5$ for item and $\Delta_u^T=10$ for user evolutions. 

\subsection{Debiasing Effectiveness Analysis (RQ4)}

\begin{figure}[h]
\vspace{-3pt}
	\centering
        \includegraphics[scale=0.38]{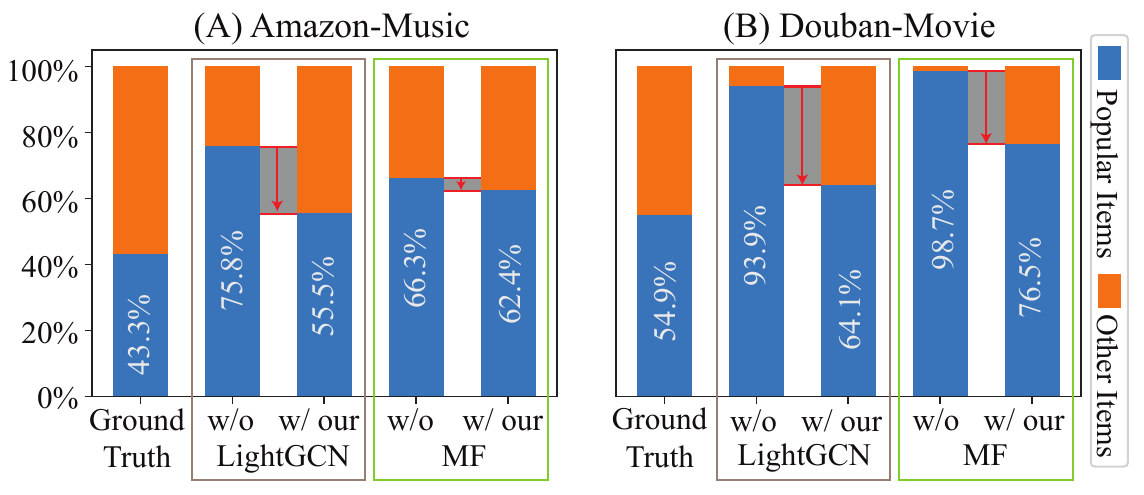}
        \vspace{-5pt}
	\caption{Recommendation frequency of popular (blue) and other (orange) items on Amazon-Music and Douban-Movie.}
    \label{fig:debiasing_effect}
    \vspace{-8pt}
\end{figure}

To explore the debiasing effectiveness of CausalEPP, we divide the items into popular and other items based on the number of interactions, \textit{i.e.,} popular items are the top 20\% of items on Amazon-Music and Douban-Movie that users frequently interact with.
As shown in Figure~\textcolor{blue}{\ref{fig:debiasing_effect}}, Amazon-Music and Douban-Movie hold a ratio of 43.3\% and 54.9\% for popular items as the ground truth, respectively.
However, the backbone models without using CausalEPP show extreme popularity bias by over-recommending popular items, \textit{e.g.,} LightGCN gives 75.8\% popular item recommendation on Amazon-Music and 93.9\% on Douban-Movie. 
CausalEPP alleviates the bias for both backbone models, \textit{e.g.,} decreasing the recommendation of popular items by LightGCN from 75.8\% to 55.5\% for Amazon-Music and 93.9\% to 64.1\% for Douban-Movie, respectively. 
Overall, CausalEPP shows significant effectiveness in mitigating popularity bias thanks to the deconfounded training and intervention mechanism.




\section{Conclusion}
In this work, our aim is to propose a popularity debiasing method that improves the recommendation performance and reduces the bias simultaneously. 
To achieve this goal, we investigate the overall issues of current studies, such as they uniformly address popularity bias for all users and ignore the evolution of both users and items. 
Our proposed debiasing algorithm measures the consistency between  {evolving personal popularity} and local popularity, disentangles quality from local popularity, and performs temporal evolution forecasting-based intervention. 
We further explored the effectiveness of key designs in our model.
{In future work, we plan to explore the applicability of our causal debiasing method in broader contexts, including mitigating biases in large language models (LLMs).}

\bibliographystyle{ACM-Reference-Format}
\bibliography{main}


\begin{thebibliography}{67}


\ifx \showCODEN    \undefined \def \showCODEN     #1{\unskip}     \fi
\ifx \showDOI      \undefined \def \showDOI       #1{#1}\fi
\ifx \showISBNx    \undefined \def \showISBNx     #1{\unskip}     \fi
\ifx \showISBNxiii \undefined \def \showISBNxiii  #1{\unskip}     \fi
\ifx \showISSN     \undefined \def \showISSN      #1{\unskip}     \fi
\ifx \showLCCN     \undefined \def \showLCCN      #1{\unskip}     \fi
\ifx \shownote     \undefined \def \shownote      #1{#1}          \fi
\ifx \showarticletitle \undefined \def \showarticletitle #1{#1}   \fi
\ifx \showURL      \undefined \def \showURL       {\relax}        \fi
\providecommand\bibfield[2]{#2}
\providecommand\bibinfo[2]{#2}
\providecommand\natexlab[1]{#1}
\providecommand\showeprint[2][]{arXiv:#2}

\bibitem[Abdollahpouri(2020)]%
        {abdollahpouri2020popularity}
\bibfield{author}{\bibinfo{person}{Himan Abdollahpouri}.} \bibinfo{year}{2020}\natexlab{}.
\newblock \emph{\bibinfo{title}{Popularity bias in recommendation: a multi-stakeholder perspective}}.
\newblock \bibinfo{thesistype}{Ph.\,D. Dissertation}. \bibinfo{school}{University of Colorado at Boulder}.
\newblock


\bibitem[Abdollahpouri et~al\mbox{.}(2020)]%
        {DBLP:conf/recsys/AbdollahpouriMB20}
\bibfield{author}{\bibinfo{person}{Himan Abdollahpouri}, \bibinfo{person}{Masoud Mansoury}, \bibinfo{person}{Robin Burke}, {and} \bibinfo{person}{Bamshad Mobasher}.} \bibinfo{year}{2020}\natexlab{}.
\newblock \showarticletitle{The Connection Between Popularity Bias, Calibration, and Fairness in Recommendation}. In \bibinfo{booktitle}{\emph{RecSys 2020: Fourteenth {ACM} Conference on Recommender Systems, Virtual Event, Brazil, September 22-26, 2020}}. \bibinfo{publisher}{{ACM}}, \bibinfo{pages}{726--731}.
\newblock


\bibitem[Anderson(2006)]%
        {anderson2006long}
\bibfield{author}{\bibinfo{person}{Chris Anderson}.} \bibinfo{year}{2006}\natexlab{}.
\newblock \bibinfo{booktitle}{\emph{The long tail: Why the future of business is selling more for less}}.
\newblock \bibinfo{publisher}{Hyperion}.
\newblock


\bibitem[Bai et~al\mbox{.}(2024)]%
        {DBLP:conf/sigir/BaiWHCHZHW24}
\bibfield{author}{\bibinfo{person}{Haoyue Bai}, \bibinfo{person}{Le Wu}, \bibinfo{person}{Min Hou}, \bibinfo{person}{Miaomiao Cai}, \bibinfo{person}{Zhuangzhuang He}, \bibinfo{person}{Yuyang Zhou}, \bibinfo{person}{Richang Hong}, {and} \bibinfo{person}{Meng Wang}.} \bibinfo{year}{2024}\natexlab{}.
\newblock \showarticletitle{Multimodality Invariant Learning for Multimedia-Based New Item Recommendation}. In \bibinfo{booktitle}{\emph{Proceedings of the 47th International {ACM} {SIGIR} Conference on Research and Development in Information Retrieval, {SIGIR} 2024, Washington DC, USA, July 14-18, 2024}}. \bibinfo{publisher}{{ACM}}, \bibinfo{pages}{677--686}.
\newblock


\bibitem[Bellog{\'{\i}}n et~al\mbox{.}(2024)]%
        {DBLP:conf/sigir/BelloginBKLMM24}
\bibfield{author}{\bibinfo{person}{Alejandro Bellog{\'{\i}}n}, \bibinfo{person}{Ludovico Boratto}, \bibinfo{person}{Styliani Kleanthous}, \bibinfo{person}{Elisabeth Lex}, \bibinfo{person}{Francesca~Maridina Malloci}, {and} \bibinfo{person}{Mirko Marras}.} \bibinfo{year}{2024}\natexlab{}.
\newblock \showarticletitle{International Workshop on Algorithmic Bias in Search and Recommendation {(BIAS)}}. In \bibinfo{booktitle}{\emph{Proceedings of the 47th International {ACM} {SIGIR} Conference on Research and Development in Information Retrieval, {SIGIR} 2024, Washington DC, USA, July 14-18, 2024}}. \bibinfo{publisher}{{ACM}}, \bibinfo{pages}{3033--3035}.
\newblock


\bibitem[Blundell(1988)]%
        {blundell1988consumer}
\bibfield{author}{\bibinfo{person}{Richard Blundell}.} \bibinfo{year}{1988}\natexlab{}.
\newblock \showarticletitle{Consumer behaviour: theory and empirical evidence—a survey}.
\newblock \bibinfo{journal}{\emph{The economic journal}} \bibinfo{volume}{98}, \bibinfo{number}{389} (\bibinfo{year}{1988}), \bibinfo{pages}{16--65}.
\newblock


\bibitem[Cai et~al\mbox{.}(2024)]%
        {DBLP:conf/kdd/Cai0WBSWZ024}
\bibfield{author}{\bibinfo{person}{Miaomiao Cai}, \bibinfo{person}{Lei Chen}, \bibinfo{person}{Yifan Wang}, \bibinfo{person}{Haoyue Bai}, \bibinfo{person}{Peijie Sun}, \bibinfo{person}{Le Wu}, \bibinfo{person}{Min Zhang}, {and} \bibinfo{person}{Meng Wang}.} \bibinfo{year}{2024}\natexlab{}.
\newblock \showarticletitle{Popularity-Aware Alignment and Contrast for Mitigating Popularity Bias}. In \bibinfo{booktitle}{\emph{Proceedings of the 30th {ACM} {SIGKDD} Conference on Knowledge Discovery and Data Mining, {KDD} 2024, Barcelona, Spain, August 25-29, 2024}}. \bibinfo{publisher}{{ACM}}, \bibinfo{pages}{187--198}.
\newblock


\bibitem[Chen et~al\mbox{.}(2023)]%
        {DBLP:journals/tois/0007D0F0023}
\bibfield{author}{\bibinfo{person}{Jiawei Chen}, \bibinfo{person}{Hande Dong}, \bibinfo{person}{Xiang Wang}, \bibinfo{person}{Fuli Feng}, \bibinfo{person}{Meng Wang}, {and} \bibinfo{person}{Xiangnan He}.} \bibinfo{year}{2023}\natexlab{}.
\newblock \showarticletitle{Bias and Debias in Recommender System: {A} Survey and Future Directions}.
\newblock \bibinfo{journal}{\emph{{ACM} Trans. Inf. Syst.}} \bibinfo{volume}{41}, \bibinfo{number}{3} (\bibinfo{year}{2023}), \bibinfo{pages}{67:1--67:39}.
\newblock


\bibitem[Chen et~al\mbox{.}(2024b)]%
        {DBLP:conf/sigir/Chen0GWWH24}
\bibfield{author}{\bibinfo{person}{Jiaju Chen}, \bibinfo{person}{Wenjie Wang}, \bibinfo{person}{Chongming Gao}, \bibinfo{person}{Peng Wu}, \bibinfo{person}{Jianxiong Wei}, {and} \bibinfo{person}{Qingsong Hua}.} \bibinfo{year}{2024}\natexlab{b}.
\newblock \showarticletitle{Treatment Effect Estimation for User Interest Exploration on Recommender Systems}. In \bibinfo{booktitle}{\emph{Proceedings of the 47th International {ACM} {SIGIR} Conference on Research and Development in Information Retrieval, {SIGIR} 2024, Washington DC, USA, July 14-18, 2024}}. \bibinfo{publisher}{{ACM}}, \bibinfo{pages}{1861--1871}.
\newblock


\bibitem[Chen et~al\mbox{.}(2024c)]%
        {DBLP:journals/fcsc/ChenWCXLH24}
\bibfield{author}{\bibinfo{person}{Jiajia Chen}, \bibinfo{person}{Jiancan Wu}, \bibinfo{person}{Jiawei Chen}, \bibinfo{person}{Xin Xin}, \bibinfo{person}{Yong Li}, {and} \bibinfo{person}{Xiangnan He}.} \bibinfo{year}{2024}\natexlab{c}.
\newblock \showarticletitle{How graph convolutions amplify popularity bias for recommendation?}
\newblock \bibinfo{journal}{\emph{Frontiers Comput. Sci.}} \bibinfo{volume}{18}, \bibinfo{number}{5} (\bibinfo{year}{2024}), \bibinfo{pages}{185603}.
\newblock


\bibitem[Chen et~al\mbox{.}(2022)]%
        {DBLP:conf/www/ChenLLMX22}
\bibfield{author}{\bibinfo{person}{Yongjun Chen}, \bibinfo{person}{Zhiwei Liu}, \bibinfo{person}{Jia Li}, \bibinfo{person}{Julian~J. McAuley}, {and} \bibinfo{person}{Caiming Xiong}.} \bibinfo{year}{2022}\natexlab{}.
\newblock \showarticletitle{Intent Contrastive Learning for Sequential Recommendation}. In \bibinfo{booktitle}{\emph{{WWW} '22: The {ACM} Web Conference 2022, Virtual Event, Lyon, France, April 25 - 29, 2022}}. \bibinfo{publisher}{{ACM}}, \bibinfo{pages}{2172--2182}.
\newblock


\bibitem[Chen et~al\mbox{.}(2024a)]%
        {DBLP:conf/kdd/0001TS0K24}
\bibfield{author}{\bibinfo{person}{Yankai Chen}, \bibinfo{person}{Quoc{-}Tuan Truong}, \bibinfo{person}{Xin Shen}, \bibinfo{person}{Jin Li}, {and} \bibinfo{person}{Irwin King}.} \bibinfo{year}{2024}\natexlab{a}.
\newblock \showarticletitle{Shopping Trajectory Representation Learning with Pre-training for E-commerce Customer Understanding and Recommendation}. In \bibinfo{booktitle}{\emph{Proceedings of the 30th {ACM} {SIGKDD} Conference on Knowledge Discovery and Data Mining, {KDD} 2024, Barcelona, Spain, August 25-29, 2024}}. \bibinfo{publisher}{{ACM}}, \bibinfo{pages}{385--396}.
\newblock


\bibitem[Ding et~al\mbox{.}(2024)]%
        {DBLP:journals/tnn/DingFHLSZ24}
\bibfield{author}{\bibinfo{person}{Sihao Ding}, \bibinfo{person}{Fuli Feng}, \bibinfo{person}{Xiangnan He}, \bibinfo{person}{Yong Liao}, \bibinfo{person}{Jun Shi}, {and} \bibinfo{person}{Yongdong Zhang}.} \bibinfo{year}{2024}\natexlab{}.
\newblock \showarticletitle{Causal Incremental Graph Convolution for Recommender System Retraining}.
\newblock \bibinfo{journal}{\emph{{IEEE} Trans. Neural Networks Learn. Syst.}} \bibinfo{volume}{35}, \bibinfo{number}{4} (\bibinfo{year}{2024}), \bibinfo{pages}{4718--4728}.
\newblock


\bibitem[Du et~al\mbox{.}(2024b)]%
        {DBLP:conf/sigir/0003Y00024}
\bibfield{author}{\bibinfo{person}{Jing Du}, \bibinfo{person}{Zesheng Ye}, \bibinfo{person}{Bin Guo}, \bibinfo{person}{Zhiwen Yu}, {and} \bibinfo{person}{Lina Yao}.} \bibinfo{year}{2024}\natexlab{b}.
\newblock \showarticletitle{Identifiability of Cross-Domain Recommendation via Causal Subspace Disentanglement}. In \bibinfo{booktitle}{\emph{Proceedings of the 47th International {ACM} {SIGIR} Conference on Research and Development in Information Retrieval, {SIGIR} 2024, Washington DC, USA, July 14-18, 2024}}. \bibinfo{publisher}{{ACM}}, \bibinfo{pages}{2091--2101}.
\newblock


\bibitem[Du et~al\mbox{.}(2024a)]%
        {DBLP:conf/kdd/DuW000024}
\bibfield{author}{\bibinfo{person}{Yingpeng Du}, \bibinfo{person}{Ziyan Wang}, \bibinfo{person}{Zhu Sun}, \bibinfo{person}{Yining Ma}, \bibinfo{person}{Hongzhi Liu}, {and} \bibinfo{person}{Jie Zhang}.} \bibinfo{year}{2024}\natexlab{a}.
\newblock \showarticletitle{Disentangled Multi-interest Representation Learning for Sequential Recommendation}. In \bibinfo{booktitle}{\emph{Proceedings of the 30th {ACM} {SIGKDD} Conference on Knowledge Discovery and Data Mining, {KDD} 2024, Barcelona, Spain, August 25-29, 2024}}. \bibinfo{publisher}{{ACM}}, \bibinfo{pages}{677--688}.
\newblock


\bibitem[Gao et~al\mbox{.}(2024b)]%
        {DBLP:journals/tois/GaoZWFHL24}
\bibfield{author}{\bibinfo{person}{Chen Gao}, \bibinfo{person}{Yu Zheng}, \bibinfo{person}{Wenjie Wang}, \bibinfo{person}{Fuli Feng}, \bibinfo{person}{Xiangnan He}, {and} \bibinfo{person}{Yong Li}.} \bibinfo{year}{2024}\natexlab{b}.
\newblock \showarticletitle{Causal Inference in Recommender Systems: {A} Survey and Future Directions}.
\newblock \bibinfo{journal}{\emph{{ACM} Trans. Inf. Syst.}} \bibinfo{volume}{42}, \bibinfo{number}{4} (\bibinfo{year}{2024}), \bibinfo{pages}{88:1--88:32}.
\newblock


\bibitem[Gao et~al\mbox{.}(2024a)]%
        {DBLP:conf/www/GaoFTYCRR24}
\bibfield{author}{\bibinfo{person}{Shen Gao}, \bibinfo{person}{Jiabao Fang}, \bibinfo{person}{Quan Tu}, \bibinfo{person}{Zhitao Yao}, \bibinfo{person}{Zhumin Chen}, \bibinfo{person}{Pengjie Ren}, {and} \bibinfo{person}{Zhaochun Ren}.} \bibinfo{year}{2024}\natexlab{a}.
\newblock \showarticletitle{Generative News Recommendation}. In \bibinfo{booktitle}{\emph{Proceedings of the {ACM} on Web Conference 2024, {WWW} 2024, Singapore, May 13-17, 2024}}. \bibinfo{publisher}{{ACM}}, \bibinfo{pages}{3444--3453}.
\newblock


\bibitem[Gao et~al\mbox{.}(2017)]%
        {gao2017bounds}
\bibfield{author}{\bibinfo{person}{Xiang Gao}, \bibinfo{person}{Meera Sitharam}, {and} \bibinfo{person}{Adrian~E Roitberg}.} \bibinfo{year}{2017}\natexlab{}.
\newblock \showarticletitle{Bounds on the Jensen gap, and implications for mean-concentrated distributions}.
\newblock \bibinfo{journal}{\emph{arXiv preprint arXiv:1712.05267}} (\bibinfo{year}{2017}).
\newblock


\bibitem[Ge et~al\mbox{.}(2020)]%
        {DBLP:conf/sigir/GeZZPSOZ20}
\bibfield{author}{\bibinfo{person}{Yingqiang Ge}, \bibinfo{person}{Shuya Zhao}, \bibinfo{person}{Honglu Zhou}, \bibinfo{person}{Changhua Pei}, \bibinfo{person}{Fei Sun}, \bibinfo{person}{Wenwu Ou}, {and} \bibinfo{person}{Yongfeng Zhang}.} \bibinfo{year}{2020}\natexlab{}.
\newblock \showarticletitle{Understanding Echo Chambers in E-commerce Recommender Systems}. In \bibinfo{booktitle}{\emph{Proceedings of the 43rd International {ACM} {SIGIR} conference on research and development in Information Retrieval, {SIGIR} 2020, Virtual Event, China, July 25-30, 2020}}. \bibinfo{publisher}{{ACM}}, \bibinfo{pages}{2261--2270}.
\newblock


\bibitem[Geng et~al\mbox{.}(2022)]%
        {DBLP:conf/recsys/Geng0FGZ22}
\bibfield{author}{\bibinfo{person}{Shijie Geng}, \bibinfo{person}{Shuchang Liu}, \bibinfo{person}{Zuohui Fu}, \bibinfo{person}{Yingqiang Ge}, {and} \bibinfo{person}{Yongfeng Zhang}.} \bibinfo{year}{2022}\natexlab{}.
\newblock \showarticletitle{Recommendation as Language Processing {(RLP):} {A} Unified Pretrain, Personalized Prompt {\&} Predict Paradigm {(P5)}}. In \bibinfo{booktitle}{\emph{RecSys '22: Sixteenth {ACM} Conference on Recommender Systems, Seattle, WA, USA, September 18 - 23, 2022}}. \bibinfo{publisher}{{ACM}}, \bibinfo{pages}{299--315}.
\newblock


\bibitem[He et~al\mbox{.}(2020)]%
        {DBLP:conf/sigir/0001DWLZ020}
\bibfield{author}{\bibinfo{person}{Xiangnan He}, \bibinfo{person}{Kuan Deng}, \bibinfo{person}{Xiang Wang}, \bibinfo{person}{Yan Li}, \bibinfo{person}{Yong{-}Dong Zhang}, {and} \bibinfo{person}{Meng Wang}.} \bibinfo{year}{2020}\natexlab{}.
\newblock \showarticletitle{LightGCN: Simplifying and Powering Graph Convolution Network for Recommendation}. In \bibinfo{booktitle}{\emph{Proceedings of the 43rd International {ACM} {SIGIR} conference on research and development in Information Retrieval, {SIGIR} 2020, Virtual Event, China, July 25-30, 2020}}. \bibinfo{publisher}{{ACM}}, \bibinfo{pages}{639--648}.
\newblock


\bibitem[Huang et~al\mbox{.}(2024)]%
        {DBLP:conf/sigir/0010OMHR24}
\bibfield{author}{\bibinfo{person}{Jin Huang}, \bibinfo{person}{Harrie Oosterhuis}, \bibinfo{person}{Masoud Mansoury}, \bibinfo{person}{Herke van Hoof}, {and} \bibinfo{person}{Maarten de Rijke}.} \bibinfo{year}{2024}\natexlab{}.
\newblock \showarticletitle{Going Beyond Popularity and Positivity Bias: Correcting for Multifactorial Bias in Recommender Systems}. In \bibinfo{booktitle}{\emph{Proceedings of the 47th International {ACM} {SIGIR} Conference on Research and Development in Information Retrieval, {SIGIR} 2024, Washington DC, USA, July 14-18, 2024}}. \bibinfo{publisher}{{ACM}}, \bibinfo{pages}{416--426}.
\newblock


\bibitem[J{\"{a}}rvelin and Kek{\"{a}}l{\"{a}}inen(2002)]%
        {DBLP:journals/tois/JarvelinK02}
\bibfield{author}{\bibinfo{person}{Kalervo J{\"{a}}rvelin} {and} \bibinfo{person}{Jaana Kek{\"{a}}l{\"{a}}inen}.} \bibinfo{year}{2002}\natexlab{}.
\newblock \showarticletitle{Cumulated gain-based evaluation of {IR} techniques}.
\newblock \bibinfo{journal}{\emph{{ACM} Trans. Inf. Syst.}} \bibinfo{volume}{20}, \bibinfo{number}{4} (\bibinfo{year}{2002}), \bibinfo{pages}{422--446}.
\newblock


\bibitem[Jiang et~al\mbox{.}(2024)]%
        {DBLP:conf/www/JiangG0CY24}
\bibfield{author}{\bibinfo{person}{Wei Jiang}, \bibinfo{person}{Xinyi Gao}, \bibinfo{person}{Guandong Xu}, \bibinfo{person}{Tong Chen}, {and} \bibinfo{person}{Hongzhi Yin}.} \bibinfo{year}{2024}\natexlab{}.
\newblock \showarticletitle{Challenging Low Homophily in Social Recommendation}. In \bibinfo{booktitle}{\emph{Proceedings of the {ACM} on Web Conference 2024, {WWW} 2024, Singapore, May 13-17, 2024}}. \bibinfo{publisher}{{ACM}}, \bibinfo{pages}{3476--3484}.
\newblock


\bibitem[Jing et~al\mbox{.}(2023)]%
        {DBLP:conf/cikm/JingZ0023}
\bibfield{author}{\bibinfo{person}{Jiazheng Jing}, \bibinfo{person}{Yinan Zhang}, \bibinfo{person}{Xin Zhou}, {and} \bibinfo{person}{Zhiqi Shen}.} \bibinfo{year}{2023}\natexlab{}.
\newblock \showarticletitle{Capturing Popularity Trends: {A} Simplistic Non-Personalized Approach for Enhanced Item Recommendation}. In \bibinfo{booktitle}{\emph{Proceedings of the 32nd {ACM} International Conference on Information and Knowledge Management, {CIKM} 2023, Birmingham, United Kingdom, October 21-25, 2023}}. \bibinfo{publisher}{{ACM}}, \bibinfo{pages}{1014--1024}.
\newblock


\bibitem[Joachims et~al\mbox{.}(2018)]%
        {DBLP:conf/ijcai/JoachimsSS18}
\bibfield{author}{\bibinfo{person}{Thorsten Joachims}, \bibinfo{person}{Adith Swaminathan}, {and} \bibinfo{person}{Tobias Schnabel}.} \bibinfo{year}{2018}\natexlab{}.
\newblock \showarticletitle{Unbiased Learning-to-Rank with Biased Feedback}. In \bibinfo{booktitle}{\emph{Proceedings of the Twenty-Seventh International Joint Conference on Artificial Intelligence, {IJCAI} 2018, July 13-19, 2018, Stockholm, Sweden}}. \bibinfo{publisher}{ijcai.org}, \bibinfo{pages}{5284--5288}.
\newblock


\bibitem[Lakkaraju et~al\mbox{.}(2013)]%
        {DBLP:conf/icwsm/LakkarajuML13}
\bibfield{author}{\bibinfo{person}{Himabindu Lakkaraju}, \bibinfo{person}{Julian~J. McAuley}, {and} \bibinfo{person}{Jure Leskovec}.} \bibinfo{year}{2013}\natexlab{}.
\newblock \showarticletitle{What's in a Name? Understanding the Interplay between Titles, Content, and Communities in Social Media}. In \bibinfo{booktitle}{\emph{Proceedings of the Seventh International Conference on Weblogs and Social Media, {ICWSM} 2013, Cambridge, Massachusetts, USA, July 8-11, 2013}}. \bibinfo{publisher}{The {AAAI} Press}.
\newblock


\bibitem[Liang et~al\mbox{.}(2016)]%
        {liang2016causal}
\bibfield{author}{\bibinfo{person}{Dawen Liang}, \bibinfo{person}{Laurent Charlin}, {and} \bibinfo{person}{David~M Blei}.} \bibinfo{year}{2016}\natexlab{}.
\newblock \showarticletitle{Causal inference for recommendation}. In \bibinfo{booktitle}{\emph{Causation: Foundation to Application, Workshop at UAI. AUAI}}, Vol.~\bibinfo{volume}{6}. \bibinfo{pages}{108}.
\newblock


\bibitem[Lin et~al\mbox{.}(2025)]%
        {DBLP:journals/corr/abs-2404-12008}
\bibfield{author}{\bibinfo{person}{Siyi Lin}, \bibinfo{person}{Chongming Gao}, \bibinfo{person}{Jiawei Chen}, \bibinfo{person}{Sheng Zhou}, \bibinfo{person}{Binbin Hu}, \bibinfo{person}{Yan Feng}, \bibinfo{person}{Chun Chen}, {and} \bibinfo{person}{Can Wang}.} \bibinfo{year}{2025}\natexlab{}.
\newblock \showarticletitle{How Do Recommendation Models Amplify Popularity Bias? An Analysis from the Spectral Perspective}. In \bibinfo{booktitle}{\emph{Proceedings of the Eighteenth ACM International Conference on Web Search and Data Mining}} \emph{(\bibinfo{series}{WSDM ’25})}. \bibinfo{publisher}{ACM}, \bibinfo{pages}{659–668}.
\newblock


\bibitem[Luo and Wu(2023)]%
        {DBLP:conf/sigir/LuoW23}
\bibfield{author}{\bibinfo{person}{Fangyuan Luo} {and} \bibinfo{person}{Jun Wu}.} \bibinfo{year}{2023}\natexlab{}.
\newblock \showarticletitle{User-Dependent Learning to Debias for Recommendation}. In \bibinfo{booktitle}{\emph{Proceedings of the 46th International {ACM} {SIGIR} Conference on Research and Development in Information Retrieval, {SIGIR} 2023, Taipei, Taiwan, July 23-27, 2023}}. \bibinfo{publisher}{{ACM}}, \bibinfo{pages}{2491--2495}.
\newblock


\bibitem[Luo et~al\mbox{.}(2024)]%
        {luo2024survey}
\bibfield{author}{\bibinfo{person}{Huishi Luo}, \bibinfo{person}{Fuzhen Zhuang}, \bibinfo{person}{Ruobing Xie}, \bibinfo{person}{Hengshu Zhu}, \bibinfo{person}{Deqing Wang}, \bibinfo{person}{Zhulin An}, {and} \bibinfo{person}{Yongjun Xu}.} \bibinfo{year}{2024}\natexlab{}.
\newblock \showarticletitle{A survey on causal inference for recommendation}.
\newblock \bibinfo{journal}{\emph{The Innovation}} (\bibinfo{year}{2024}).
\newblock


\bibitem[M{\"o}ller et~al\mbox{.}(2020)]%
        {moller2020not}
\bibfield{author}{\bibinfo{person}{Judith M{\"o}ller}, \bibinfo{person}{Damian Trilling}, \bibinfo{person}{Natali Helberger}, {and} \bibinfo{person}{Bram van Es}.} \bibinfo{year}{2020}\natexlab{}.
\newblock \showarticletitle{Do not blame it on the algorithm: an empirical assessment of multiple recommender systems and their impact on content diversity}.
\newblock In \bibinfo{booktitle}{\emph{Digital media, political polarization and challenges to democracy}}. \bibinfo{publisher}{Routledge}, \bibinfo{pages}{45--63}.
\newblock


\bibitem[Nayak et~al\mbox{.}(2023)]%
        {DBLP:conf/sigir/NayakGM23}
\bibfield{author}{\bibinfo{person}{Ashutosh Nayak}, \bibinfo{person}{Mayur Garg}, {and} \bibinfo{person}{Rajasekhara Reddy~Duvvuru Muni}.} \bibinfo{year}{2023}\natexlab{}.
\newblock \showarticletitle{News Popularity Beyond the Click-Through-Rate for Personalized Recommendations}. In \bibinfo{booktitle}{\emph{Proceedings of the 46th International {ACM} {SIGIR} Conference on Research and Development in Information Retrieval, {SIGIR} 2023, Taipei, Taiwan, July 23-27, 2023}}. \bibinfo{publisher}{{ACM}}, \bibinfo{pages}{1396--1405}.
\newblock


\bibitem[Nguyen et~al\mbox{.}(2014)]%
        {DBLP:conf/www/NguyenHHTK14}
\bibfield{author}{\bibinfo{person}{Tien~T. Nguyen}, \bibinfo{person}{Pik{-}Mai Hui}, \bibinfo{person}{F.~Maxwell Harper}, \bibinfo{person}{Loren~G. Terveen}, {and} \bibinfo{person}{Joseph~A. Konstan}.} \bibinfo{year}{2014}\natexlab{}.
\newblock \showarticletitle{Exploring the filter bubble: the effect of using recommender systems on content diversity}. In \bibinfo{booktitle}{\emph{23rd International World Wide Web Conference, {WWW} '14, Seoul, Republic of Korea, April 7-11, 2014}}. \bibinfo{publisher}{{ACM}}, \bibinfo{pages}{677--686}.
\newblock


\bibitem[Ning et~al\mbox{.}(2024)]%
        {DBLP:conf/www/NingC0KHH024}
\bibfield{author}{\bibinfo{person}{Wentao Ning}, \bibinfo{person}{Reynold Cheng}, \bibinfo{person}{Xiao Yan}, \bibinfo{person}{Ben Kao}, \bibinfo{person}{Nan Huo}, \bibinfo{person}{Nur Al~Hasan Haldar}, {and} \bibinfo{person}{Bo Tang}.} \bibinfo{year}{2024}\natexlab{}.
\newblock \showarticletitle{Debiasing Recommendation with Personal Popularity}. In \bibinfo{booktitle}{\emph{Proceedings of the {ACM} on Web Conference 2024, {WWW} 2024, Singapore, May 13-17, 2024}}. \bibinfo{publisher}{{ACM}}, \bibinfo{pages}{3400--3409}.
\newblock


\bibitem[Pan et~al\mbox{.}(2024)]%
        {DBLP:conf/kdd/PanXWYLQQLX024}
\bibfield{author}{\bibinfo{person}{Junwei Pan}, \bibinfo{person}{Wei Xue}, \bibinfo{person}{Ximei Wang}, \bibinfo{person}{Haibin Yu}, \bibinfo{person}{Xun Liu}, \bibinfo{person}{Shijie Quan}, \bibinfo{person}{Xueming Qiu}, \bibinfo{person}{Dapeng Liu}, \bibinfo{person}{Lei Xiao}, {and} \bibinfo{person}{Jie Jiang}.} \bibinfo{year}{2024}\natexlab{}.
\newblock \showarticletitle{Ads Recommendation in a Collapsed and Entangled World}. In \bibinfo{booktitle}{\emph{Proceedings of the 30th {ACM} {SIGKDD} Conference on Knowledge Discovery and Data Mining, {KDD} 2024, Barcelona, Spain, August 25-29, 2024}}. \bibinfo{publisher}{{ACM}}, \bibinfo{pages}{5566--5577}.
\newblock


\bibitem[Pearl(2009)]%
        {pearl2009causality}
\bibfield{author}{\bibinfo{person}{J Pearl}.} \bibinfo{year}{2009}\natexlab{}.
\newblock \bibinfo{booktitle}{\emph{Causality}}.
\newblock \bibinfo{publisher}{Cambridge university press}.
\newblock


\bibitem[Rendle et~al\mbox{.}(2009)]%
        {DBLP:conf/uai/RendleFGS09}
\bibfield{author}{\bibinfo{person}{Steffen Rendle}, \bibinfo{person}{Christoph Freudenthaler}, \bibinfo{person}{Zeno Gantner}, {and} \bibinfo{person}{Lars Schmidt{-}Thieme}.} \bibinfo{year}{2009}\natexlab{}.
\newblock \showarticletitle{{BPR:} Bayesian Personalized Ranking from Implicit Feedback}. In \bibinfo{booktitle}{\emph{{UAI} 2009, Proceedings of the Twenty-Fifth Conference on Uncertainty in Artificial Intelligence, Montreal, QC, Canada, June 18-21, 2009}}. \bibinfo{publisher}{{AUAI} Press}, \bibinfo{pages}{452--461}.
\newblock


\bibitem[Saito et~al\mbox{.}(2020)]%
        {DBLP:conf/wsdm/SaitoYNSN20}
\bibfield{author}{\bibinfo{person}{Yuta Saito}, \bibinfo{person}{Suguru Yaginuma}, \bibinfo{person}{Yuta Nishino}, \bibinfo{person}{Hayato Sakata}, {and} \bibinfo{person}{Kazuhide Nakata}.} \bibinfo{year}{2020}\natexlab{}.
\newblock \showarticletitle{Unbiased Recommender Learning from Missing-Not-At-Random Implicit Feedback}. In \bibinfo{booktitle}{\emph{{WSDM} '20: The Thirteenth {ACM} International Conference on Web Search and Data Mining, Houston, TX, USA, February 3-7, 2020}}. \bibinfo{publisher}{{ACM}}, \bibinfo{pages}{501--509}.
\newblock


\bibitem[Sanders(1987)]%
        {sanders1987pareto}
\bibfield{author}{\bibinfo{person}{Robert Sanders}.} \bibinfo{year}{1987}\natexlab{}.
\newblock \showarticletitle{The Pareto principle: its use and abuse}.
\newblock \bibinfo{journal}{\emph{Journal of Services Marketing}} \bibinfo{volume}{1}, \bibinfo{number}{2} (\bibinfo{year}{1987}), \bibinfo{pages}{37--40}.
\newblock


\bibitem[Schnabel et~al\mbox{.}(2016)]%
        {DBLP:conf/icml/SchnabelSSCJ16}
\bibfield{author}{\bibinfo{person}{Tobias Schnabel}, \bibinfo{person}{Adith Swaminathan}, \bibinfo{person}{Ashudeep Singh}, \bibinfo{person}{Navin Chandak}, {and} \bibinfo{person}{Thorsten Joachims}.} \bibinfo{year}{2016}\natexlab{}.
\newblock \showarticletitle{Recommendations as Treatments: Debiasing Learning and Evaluation}. In \bibinfo{booktitle}{\emph{Proceedings of the 33nd International Conference on Machine Learning, {ICML} 2016, New York City, NY, USA, June 19-24, 2016}} \emph{(\bibinfo{series}{{JMLR} Workshop and Conference Proceedings}, Vol.~\bibinfo{volume}{48})}. \bibinfo{publisher}{JMLR.org}, \bibinfo{pages}{1670--1679}.
\newblock


\bibitem[Shang et~al\mbox{.}(2024)]%
        {YongLi-WWW24}
\bibfield{author}{\bibinfo{person}{Yu Shang}, \bibinfo{person}{Chen Gao}, \bibinfo{person}{Jiansheng Chen}, \bibinfo{person}{Depeng Jin}, {and} \bibinfo{person}{Yong Li}.} \bibinfo{year}{2024}\natexlab{}.
\newblock \showarticletitle{Improving Item-side Fairness of Multimodal Recommendation via Modality Debiasing}. In \bibinfo{booktitle}{\emph{Proceedings of the ACM Web Conference 2024}}. \bibinfo{pages}{4697–4705}.
\newblock
\showISBNx{9798400701719}


\bibitem[Shi et~al\mbox{.}(2024)]%
        {DBLP:conf/sigir/ShiZZF024}
\bibfield{author}{\bibinfo{person}{Tianhao Shi}, \bibinfo{person}{Yang Zhang}, \bibinfo{person}{Jizhi Zhang}, \bibinfo{person}{Fuli Feng}, {and} \bibinfo{person}{Xiangnan He}.} \bibinfo{year}{2024}\natexlab{}.
\newblock \showarticletitle{Fair Recommendations with Limited Sensitive Attributes: {A} Distributionally Robust Optimization Approach}. In \bibinfo{booktitle}{\emph{Proceedings of the 47th International {ACM} {SIGIR} Conference on Research and Development in Information Retrieval, {SIGIR} 2024, Washington DC, USA, July 14-18, 2024}}. \bibinfo{publisher}{{ACM}}, \bibinfo{pages}{448--457}.
\newblock


\bibitem[Song et~al\mbox{.}(2019)]%
        {DBLP:conf/wsdm/Song0WCZT19}
\bibfield{author}{\bibinfo{person}{Weiping Song}, \bibinfo{person}{Zhiping Xiao}, \bibinfo{person}{Yifan Wang}, \bibinfo{person}{Laurent Charlin}, \bibinfo{person}{Ming Zhang}, {and} \bibinfo{person}{Jian Tang}.} \bibinfo{year}{2019}\natexlab{}.
\newblock \showarticletitle{Session-Based Social Recommendation via Dynamic Graph Attention Networks}. In \bibinfo{booktitle}{\emph{Proceedings of the Twelfth {ACM} International Conference on Web Search and Data Mining, {WSDM} 2019, Melbourne, VIC, Australia, February 11-15, 2019}}. \bibinfo{publisher}{{ACM}}, \bibinfo{pages}{555--563}.
\newblock


\bibitem[Sun et~al\mbox{.}(2019)]%
        {DBLP:conf/cikm/SunLWPLOJ19}
\bibfield{author}{\bibinfo{person}{Fei Sun}, \bibinfo{person}{Jun Liu}, \bibinfo{person}{Jian Wu}, \bibinfo{person}{Changhua Pei}, \bibinfo{person}{Xiao Lin}, \bibinfo{person}{Wenwu Ou}, {and} \bibinfo{person}{Peng Jiang}.} \bibinfo{year}{2019}\natexlab{}.
\newblock \showarticletitle{BERT4Rec: Sequential Recommendation with Bidirectional Encoder Representations from Transformer}. In \bibinfo{booktitle}{\emph{Proceedings of the 28th {ACM} International Conference on Information and Knowledge Management, {CIKM} 2019, Beijing, China, November 3-7, 2019}}. \bibinfo{publisher}{{ACM}}, \bibinfo{pages}{1441--1450}.
\newblock


\bibitem[Tang et~al\mbox{.}(2023)]%
        {DBLP:conf/kdd/Tang0WGXZ0MZ23}
\bibfield{author}{\bibinfo{person}{Shisong Tang}, \bibinfo{person}{Qing Li}, \bibinfo{person}{Dingmin Wang}, \bibinfo{person}{Ci Gao}, \bibinfo{person}{Wentao Xiao}, \bibinfo{person}{Dan Zhao}, \bibinfo{person}{Yong Jiang}, \bibinfo{person}{Qian Ma}, {and} \bibinfo{person}{Aoyang Zhang}.} \bibinfo{year}{2023}\natexlab{}.
\newblock \showarticletitle{Counterfactual Video Recommendation for Duration Debiasing}. In \bibinfo{booktitle}{\emph{Proceedings of the 29th {ACM} {SIGKDD} Conference on Knowledge Discovery and Data Mining, {KDD} 2023, Long Beach, CA, USA, August 6-10, 2023}}. \bibinfo{publisher}{{ACM}}, \bibinfo{pages}{4894--4903}.
\newblock


\bibitem[Wang et~al\mbox{.}(2025)]%
        {DBLP:journals/fcsc/WangLCYTZY25}
\bibfield{author}{\bibinfo{person}{Hangyu Wang}, \bibinfo{person}{Jianghao Lin}, \bibinfo{person}{Bo Chen}, \bibinfo{person}{Yang Yang}, \bibinfo{person}{Ruiming Tang}, \bibinfo{person}{Weinan Zhang}, {and} \bibinfo{person}{Yong Yu}.} \bibinfo{year}{2025}\natexlab{}.
\newblock \showarticletitle{Towards efficient and effective unlearning of large language models for recommendation}.
\newblock \bibinfo{journal}{\emph{Frontiers Comput. Sci.}} \bibinfo{volume}{19}, \bibinfo{number}{3} (\bibinfo{year}{2025}), \bibinfo{pages}{193327}.
\newblock


\bibitem[Wang et~al\mbox{.}(2018a)]%
        {DBLP:conf/kdd/WangHZZZL18}
\bibfield{author}{\bibinfo{person}{Jizhe Wang}, \bibinfo{person}{Pipei Huang}, \bibinfo{person}{Huan Zhao}, \bibinfo{person}{Zhibo Zhang}, \bibinfo{person}{Binqiang Zhao}, {and} \bibinfo{person}{Dik~Lun Lee}.} \bibinfo{year}{2018}\natexlab{a}.
\newblock \showarticletitle{Billion-scale Commodity Embedding for E-commerce Recommendation in Alibaba}. In \bibinfo{booktitle}{\emph{Proceedings of the 24th {ACM} {SIGKDD} International Conference on Knowledge Discovery {\&} Data Mining, {KDD} 2018, London, UK, August 19-23, 2018}}. \bibinfo{publisher}{{ACM}}, \bibinfo{pages}{839--848}.
\newblock


\bibitem[Wang et~al\mbox{.}(2021)]%
        {DBLP:conf/kdd/WangF0WC21}
\bibfield{author}{\bibinfo{person}{Wenjie Wang}, \bibinfo{person}{Fuli Feng}, \bibinfo{person}{Xiangnan He}, \bibinfo{person}{Xiang Wang}, {and} \bibinfo{person}{Tat{-}Seng Chua}.} \bibinfo{year}{2021}\natexlab{}.
\newblock \showarticletitle{Deconfounded Recommendation for Alleviating Bias Amplification}. In \bibinfo{booktitle}{\emph{{KDD} '21: The 27th {ACM} {SIGKDD} Conference on Knowledge Discovery and Data Mining, Virtual Event, Singapore, August 14-18, 2021}}. \bibinfo{publisher}{{ACM}}, \bibinfo{pages}{1717--1725}.
\newblock


\bibitem[Wang et~al\mbox{.}(2023)]%
        {DBLP:journals/tkde/WangLYCWX23}
\bibfield{author}{\bibinfo{person}{Xiangmeng Wang}, \bibinfo{person}{Qian Li}, \bibinfo{person}{Dianer Yu}, \bibinfo{person}{Peng Cui}, \bibinfo{person}{Zhichao Wang}, {and} \bibinfo{person}{Guandong Xu}.} \bibinfo{year}{2023}\natexlab{}.
\newblock \showarticletitle{Causal Disentanglement for Semantic-Aware Intent Learning in Recommendation}.
\newblock \bibinfo{journal}{\emph{{IEEE} Trans. Knowl. Data Eng.}} \bibinfo{volume}{35}, \bibinfo{number}{10} (\bibinfo{year}{2023}), \bibinfo{pages}{9836--9849}.
\newblock


\bibitem[Wang et~al\mbox{.}(2018b)]%
        {wang2018deconfounded}
\bibfield{author}{\bibinfo{person}{Yixin Wang}, \bibinfo{person}{Dawen Liang}, \bibinfo{person}{Laurent Charlin}, {and} \bibinfo{person}{David~M Blei}.} \bibinfo{year}{2018}\natexlab{b}.
\newblock \showarticletitle{The deconfounded recommender: A causal inference approach to recommendation}.
\newblock \bibinfo{journal}{\emph{arXiv preprint arXiv:1808.06581}} (\bibinfo{year}{2018}).
\newblock


\bibitem[Wei et~al\mbox{.}(2021)]%
        {DBLP:conf/kdd/WeiFCWYH21}
\bibfield{author}{\bibinfo{person}{Tianxin Wei}, \bibinfo{person}{Fuli Feng}, \bibinfo{person}{Jiawei Chen}, \bibinfo{person}{Ziwei Wu}, \bibinfo{person}{Jinfeng Yi}, {and} \bibinfo{person}{Xiangnan He}.} \bibinfo{year}{2021}\natexlab{}.
\newblock \showarticletitle{Model-Agnostic Counterfactual Reasoning for Eliminating Popularity Bias in Recommender System}. In \bibinfo{booktitle}{\emph{{KDD} '21: The 27th {ACM} {SIGKDD} Conference on Knowledge Discovery and Data Mining, Virtual Event, Singapore, August 14-18, 2021}}. \bibinfo{publisher}{{ACM}}, \bibinfo{pages}{1791--1800}.
\newblock


\bibitem[Wei et~al\mbox{.}(2024)]%
        {DBLP:conf/www/WeiTXJH24}
\bibfield{author}{\bibinfo{person}{Wei Wei}, \bibinfo{person}{Jiabin Tang}, \bibinfo{person}{Lianghao Xia}, \bibinfo{person}{Yangqin Jiang}, {and} \bibinfo{person}{Chao Huang}.} \bibinfo{year}{2024}\natexlab{}.
\newblock \showarticletitle{PromptMM: Multi-Modal Knowledge Distillation for Recommendation with Prompt-Tuning}. In \bibinfo{booktitle}{\emph{Proceedings of the {ACM} on Web Conference 2024, {WWW} 2024, Singapore, May 13-17, 2024}}. \bibinfo{publisher}{{ACM}}, \bibinfo{pages}{3217--3228}.
\newblock


\bibitem[Wu et~al\mbox{.}(2024)]%
        {DBLP:journals/www/WuZQWGSQZZLXC24}
\bibfield{author}{\bibinfo{person}{Likang Wu}, \bibinfo{person}{Zhi Zheng}, \bibinfo{person}{Zhaopeng Qiu}, \bibinfo{person}{Hao Wang}, \bibinfo{person}{Hongchao Gu}, \bibinfo{person}{Tingjia Shen}, \bibinfo{person}{Chuan Qin}, \bibinfo{person}{Chen Zhu}, \bibinfo{person}{Hengshu Zhu}, \bibinfo{person}{Qi Liu}, \bibinfo{person}{Hui Xiong}, {and} \bibinfo{person}{Enhong Chen}.} \bibinfo{year}{2024}\natexlab{}.
\newblock \showarticletitle{A survey on large language models for recommendation}.
\newblock \bibinfo{journal}{\emph{World Wide Web {(WWW)}}} \bibinfo{volume}{27}, \bibinfo{number}{5} (\bibinfo{year}{2024}), \bibinfo{pages}{60}.
\newblock


\bibitem[Xu et~al\mbox{.}(2024)]%
        {DBLP:conf/sigir/0001DQZXXXD24}
\bibfield{author}{\bibinfo{person}{Xiaolong Xu}, \bibinfo{person}{Hongsheng Dong}, \bibinfo{person}{Lianyong Qi}, \bibinfo{person}{Xuyun Zhang}, \bibinfo{person}{Haolong Xiang}, \bibinfo{person}{Xiaoyu Xia}, \bibinfo{person}{Yanwei Xu}, {and} \bibinfo{person}{Wanchun Dou}.} \bibinfo{year}{2024}\natexlab{}.
\newblock \showarticletitle{CMCLRec: Cross-modal Contrastive Learning for User Cold-start Sequential Recommendation}. In \bibinfo{booktitle}{\emph{Proceedings of the 47th International {ACM} {SIGIR} Conference on Research and Development in Information Retrieval, {SIGIR} 2024, Washington DC, USA, July 14-18, 2024}}. \bibinfo{publisher}{{ACM}}, \bibinfo{pages}{1589--1598}.
\newblock


\bibitem[Yang et~al\mbox{.}(2024)]%
        {DBLP:conf/kdd/YangDHZXSZ24}
\bibfield{author}{\bibinfo{person}{Chen Yang}, \bibinfo{person}{Sunhao Dai}, \bibinfo{person}{Yupeng Hou}, \bibinfo{person}{Wayne~Xin Zhao}, \bibinfo{person}{Jun Xu}, \bibinfo{person}{Yang Song}, {and} \bibinfo{person}{Hengshu Zhu}.} \bibinfo{year}{2024}\natexlab{}.
\newblock \showarticletitle{Revisiting Reciprocal Recommender Systems: Metrics, Formulation, and Method}. In \bibinfo{booktitle}{\emph{Proceedings of the 30th {ACM} {SIGKDD} Conference on Knowledge Discovery and Data Mining, {KDD} 2024, Barcelona, Spain, August 25-29, 2024}}. \bibinfo{publisher}{{ACM}}, \bibinfo{pages}{3714--3723}.
\newblock


\bibitem[Yang et~al\mbox{.}(2023)]%
        {DBLP:conf/www/YangHXHLL23}
\bibfield{author}{\bibinfo{person}{Yuhao Yang}, \bibinfo{person}{Chao Huang}, \bibinfo{person}{Lianghao Xia}, \bibinfo{person}{Chunzhen Huang}, \bibinfo{person}{Da Luo}, {and} \bibinfo{person}{Kangyi Lin}.} \bibinfo{year}{2023}\natexlab{}.
\newblock \showarticletitle{Debiased Contrastive Learning for Sequential Recommendation}. In \bibinfo{booktitle}{\emph{Proceedings of the {ACM} Web Conference 2023, {WWW} 2023, Austin, TX, USA, 30 April 2023 - 4 May 2023}}. \bibinfo{publisher}{{ACM}}, \bibinfo{pages}{1063--1073}.
\newblock


\bibitem[Zhang et~al\mbox{.}(2024)]%
        {Zhang_2024}
\bibfield{author}{\bibinfo{person}{An Zhang}, \bibinfo{person}{Wenchang Ma}, \bibinfo{person}{Jingnan Zheng}, \bibinfo{person}{Xiang Wang}, {and} \bibinfo{person}{Tat-Seng Chua}.} \bibinfo{year}{2024}\natexlab{}.
\newblock \showarticletitle{Robust Collaborative Filtering to Popularity Distribution Shift}.
\newblock \bibinfo{journal}{\emph{ACM Transactions on Information Systems}} \bibinfo{volume}{42}, \bibinfo{number}{3} (\bibinfo{date}{Jan.} \bibinfo{year}{2024}), \bibinfo{pages}{1–25}.
\newblock
\showISSN{1558-2868}


\bibitem[Zhang et~al\mbox{.}(2023)]%
        {DBLP:conf/www/0003ZWYC23}
\bibfield{author}{\bibinfo{person}{An Zhang}, \bibinfo{person}{Jingnan Zheng}, \bibinfo{person}{Xiang Wang}, \bibinfo{person}{Yancheng Yuan}, {and} \bibinfo{person}{Tat{-}Seng Chua}.} \bibinfo{year}{2023}\natexlab{}.
\newblock \showarticletitle{Invariant Collaborative Filtering to Popularity Distribution Shift}. In \bibinfo{booktitle}{\emph{Proceedings of the {ACM} Web Conference 2023, {WWW} 2023, Austin, TX, USA, 30 April 2023 - 4 May 2023}}, \bibfield{editor}{\bibinfo{person}{Ying Ding}, \bibinfo{person}{Jie Tang}, \bibinfo{person}{Juan~F. Sequeda}, \bibinfo{person}{Lora Aroyo}, \bibinfo{person}{Carlos Castillo}, {and} \bibinfo{person}{Geert{-}Jan Houben}} (Eds.). \bibinfo{publisher}{{ACM}}, \bibinfo{pages}{1240--1251}.
\newblock


\bibitem[Zhang et~al\mbox{.}(2021)]%
        {DBLP:conf/sigir/ZhangF0WSL021}
\bibfield{author}{\bibinfo{person}{Yang Zhang}, \bibinfo{person}{Fuli Feng}, \bibinfo{person}{Xiangnan He}, \bibinfo{person}{Tianxin Wei}, \bibinfo{person}{Chonggang Song}, \bibinfo{person}{Guohui Ling}, {and} \bibinfo{person}{Yongdong Zhang}.} \bibinfo{year}{2021}\natexlab{}.
\newblock \showarticletitle{Causal Intervention for Leveraging Popularity Bias in Recommendation}. In \bibinfo{booktitle}{\emph{{SIGIR} '21: The 44th International {ACM} {SIGIR} Conference on Research and Development in Information Retrieval, Virtual Event, Canada, July 11-15, 2021}}. \bibinfo{publisher}{{ACM}}, \bibinfo{pages}{11--20}.
\newblock


\bibitem[Zhao et~al\mbox{.}(2022)]%
        {DBLP:conf/sigir/0002WLCZDWSLW22}
\bibfield{author}{\bibinfo{person}{Minghao Zhao}, \bibinfo{person}{Le Wu}, \bibinfo{person}{Yile Liang}, \bibinfo{person}{Lei Chen}, \bibinfo{person}{Jian Zhang}, \bibinfo{person}{Qilin Deng}, \bibinfo{person}{Kai Wang}, \bibinfo{person}{Xudong Shen}, \bibinfo{person}{Tangjie Lv}, {and} \bibinfo{person}{Runze Wu}.} \bibinfo{year}{2022}\natexlab{}.
\newblock \showarticletitle{Investigating Accuracy-Novelty Performance for Graph-based Collaborative Filtering}. In \bibinfo{booktitle}{\emph{{SIGIR} '22: The 45th International {ACM} {SIGIR} Conference on Research and Development in Information Retrieval, Madrid, Spain, July 11 - 15, 2022}}. \bibinfo{publisher}{{ACM}}, \bibinfo{pages}{50--59}.
\newblock


\bibitem[Zhao et~al\mbox{.}(2023)]%
        {DBLP:journals/tkde/ZhaoCZHCZW23}
\bibfield{author}{\bibinfo{person}{Zihao Zhao}, \bibinfo{person}{Jiawei Chen}, \bibinfo{person}{Sheng Zhou}, \bibinfo{person}{Xiangnan He}, \bibinfo{person}{Xuezhi Cao}, \bibinfo{person}{Fuzheng Zhang}, {and} \bibinfo{person}{Wei Wu}.} \bibinfo{year}{2023}\natexlab{}.
\newblock \showarticletitle{Popularity Bias is not Always Evil: Disentangling Benign and Harmful Bias for Recommendation}.
\newblock \bibinfo{journal}{\emph{{IEEE} Trans. Knowl. Data Eng.}} \bibinfo{volume}{35}, \bibinfo{number}{10} (\bibinfo{year}{2023}), \bibinfo{pages}{9920--9931}.
\newblock


\bibitem[Zheng et~al\mbox{.}(2022)]%
        {DBLP:conf/www/ZhengGCNSJL22}
\bibfield{author}{\bibinfo{person}{Yu Zheng}, \bibinfo{person}{Chen Gao}, \bibinfo{person}{Jianxin Chang}, \bibinfo{person}{Yanan Niu}, \bibinfo{person}{Yang Song}, \bibinfo{person}{Depeng Jin}, {and} \bibinfo{person}{Yong Li}.} \bibinfo{year}{2022}\natexlab{}.
\newblock \showarticletitle{Disentangling Long and Short-Term Interests for Recommendation}. In \bibinfo{booktitle}{\emph{{WWW} '22: The {ACM} Web Conference 2022, Virtual Event, Lyon, France, April 25 - 29, 2022}}. \bibinfo{publisher}{{ACM}}, \bibinfo{pages}{2256--2267}.
\newblock


\bibitem[Zheng et~al\mbox{.}(2021)]%
        {DBLP:conf/www/ZhengGLHLJ21}
\bibfield{author}{\bibinfo{person}{Yu Zheng}, \bibinfo{person}{Chen Gao}, \bibinfo{person}{Xiang Li}, \bibinfo{person}{Xiangnan He}, \bibinfo{person}{Yong Li}, {and} \bibinfo{person}{Depeng Jin}.} \bibinfo{year}{2021}\natexlab{}.
\newblock \showarticletitle{Disentangling User Interest and Conformity for Recommendation with Causal Embedding}. In \bibinfo{booktitle}{\emph{{WWW} '21: The Web Conference 2021, Virtual Event / Ljubljana, Slovenia, April 19-23, 2021}}. \bibinfo{publisher}{{ACM} / {IW3C2}}, \bibinfo{pages}{2980--2991}.
\newblock


\bibitem[Zhou et~al\mbox{.}(2020)]%
        {DBLP:conf/cikm/ZhouWZZWZWW20}
\bibfield{author}{\bibinfo{person}{Kun Zhou}, \bibinfo{person}{Hui Wang}, \bibinfo{person}{Wayne~Xin Zhao}, \bibinfo{person}{Yutao Zhu}, \bibinfo{person}{Sirui Wang}, \bibinfo{person}{Fuzheng Zhang}, \bibinfo{person}{Zhongyuan Wang}, {and} \bibinfo{person}{Ji{-}Rong Wen}.} \bibinfo{year}{2020}\natexlab{}.
\newblock \showarticletitle{S3-Rec: Self-Supervised Learning for Sequential Recommendation with Mutual Information Maximization}. In \bibinfo{booktitle}{\emph{{CIKM} '20: The 29th {ACM} International Conference on Information and Knowledge Management, Virtual Event, Ireland, October 19-23, 2020}}. \bibinfo{publisher}{{ACM}}, \bibinfo{pages}{1893--1902}.
\newblock


\bibitem[Zhu et~al\mbox{.}(2024)]%
        {DBLP:journals/tkde/ZhuZFYWH24}
\bibfield{author}{\bibinfo{person}{Xinyuan Zhu}, \bibinfo{person}{Yang Zhang}, \bibinfo{person}{Fuli Feng}, \bibinfo{person}{Xun Yang}, \bibinfo{person}{Dingxian Wang}, {and} \bibinfo{person}{Xiangnan He}.} \bibinfo{year}{2024}\natexlab{}.
\newblock \showarticletitle{Mitigating Hidden Confounding Effects for Causal Recommendation}.
\newblock \bibinfo{journal}{\emph{{IEEE} Trans. Knowl. Data Eng.}} \bibinfo{volume}{36}, \bibinfo{number}{9} (\bibinfo{year}{2024}), \bibinfo{pages}{4794--4805}.
\newblock


\bibitem[Zhu et~al\mbox{.}(2021)]%
        {DBLP:conf/wsdm/Zhu0ZZWC21}
\bibfield{author}{\bibinfo{person}{Ziwei Zhu}, \bibinfo{person}{Yun He}, \bibinfo{person}{Xing Zhao}, \bibinfo{person}{Yin Zhang}, \bibinfo{person}{Jianling Wang}, {and} \bibinfo{person}{James Caverlee}.} \bibinfo{year}{2021}\natexlab{}.
\newblock \showarticletitle{Popularity-Opportunity Bias in Collaborative Filtering}. In \bibinfo{booktitle}{\emph{{WSDM} '21, The Fourteenth {ACM} International Conference on Web Search and Data Mining, Virtual Event, Israel, March 8-12, 2021}}. \bibinfo{publisher}{{ACM}}, \bibinfo{pages}{85--93}.
\newblock


\end{thebibliography}



\end{document}